\documentclass[12pt]{article}
\usepackage{graphicx}
\usepackage{epstopdf}
\usepackage{amsthm}
\usepackage{amsmath}
\usepackage{amssymb}
\usepackage{amstext}
\usepackage{amsfonts}
\usepackage{enumerate}
\usepackage[authoryear,nonamebreak,bibstyle]{natbib}
\usepackage{footnote}
\makesavenoteenv{tabular}
\makesavenoteenv{table}
\usepackage[onehalfspacing]{setspace}
\usepackage[shortlabels]{enumitem}
\usepackage{multirow}
\usepackage[title]{appendix}
\usepackage{xcolor}
\usepackage{cancel}

\begin{document}
\baselineskip=.22in\parindent=30pt

\newtheorem{tm}{Theorem}
\newtheorem{dfn}{Definition}
\newtheorem{lma}{Lemma}
\newtheorem{assu}{Assumptions}
\newtheorem{prop}{Proposition}
\newtheorem{cro}{Corollary}
\newtheorem*{theorem*}{Theorem}
\newtheorem{example}{Example}
\newtheorem{observation}{Observation}
\newcommand{\exm}{\begin{example}}
\newcommand{\exmm}{\end{example}}
\newcommand{\obs}{\begin{observation}}
\newcommand{\obss}{\end{observation}}
\newcommand{\cor}{\begin{cro}}
\newcommand{\corr}{\end{cro}}
\newtheorem{exa}{Example}
\newcommand{\ex}{\begin{exa}}
\newcommand{\exx}{\end{exa}}
\newtheorem{remak}{Remark}
\newcommand{\rmk}{\begin{remak}}
\newcommand{\rmkk}{\end{remak}}
\newcommand{\thm}{\begin{tm}}
\newcommand{\nt}{\noindent}
\newcommand{\thmm}{\end{tm}}
\newcommand{\lm}{\begin{lma}}
\newcommand{\lmm}{\end{lma}}
\newcommand{\ass}{\begin{assu}}
\newcommand{\asss}{\end{assu}}
\newcommand{\df}{\begin{dfn}  }
\newcommand{\dff}{\end{dfn}}
\newcommand{\prp}{\begin{prop}}
\newcommand{\prpp}{\end{prop}}
\newcommand{\bqu}{\sloppy \small \begin{quote}}
\newcommand{\equ}{\end{quote} \sloppy \large}
\newcommand\cites[1]{\citeauthor{#1}'s\ (\citeyear{#1})}

\newcommand{\eq}{\begin{equation}}
\newcommand{\eqq}{\end{equation}}
\newtheorem{claim}{\it Claim}
\newcommand{\cl}{\begin{claim}}
\newcommand{\cll}{\end{claim}}
\newcommand{\bit}{\begin{itemize}}
\newcommand{\eit}{\end{itemize}}
\newcommand{\ben}{\begin{enumerate}}
\newcommand{\een}{\end{enumerate}}
\newcommand{\bcen}{\begin{center}}
\newcommand{\ecen}{\end{center}}
\newcommand{\fn}{\footnote}
\newcommand{\ds}{\begin{description}}
\newcommand{\dss}{\end{description}}
\newcommand{\prf}{\begin{proof}}
\newcommand{\prff}{\end{proof}}
\newcommand{\cs}{\begin{cases}}
\newcommand{\css}{\end{cases}}
\newcommand{\ml}{\item}
\newcommand{\lb}{\label}
\newcommand{\ra}{\rightarrow}
\newcommand{\tra}{\twoheadrightarrow}
\newcommand*{\supp}{\operatornamewithlimits{sup}\limits}
\newcommand*{\inff}{\operatornamewithlimits{inf}\limits}
\newcommand{\nf}{\normalfont}
\renewcommand{\Re}{\mathbb{R}}
\newcommand*{\mmax}{\operatornamewithlimits{max}\limits}
\newcommand*{\mmin}{\operatornamewithlimits{min}\limits}
\newcommand*{\argmax}{\operatornamewithlimits{arg max}\limits}
\newcommand*{\argmin}{\operatornamewithlimits{arg min}\limits}

\newcommand{\CR}{\mathcal R}
\newcommand{\CC}{\mathcal C}
\newcommand{\CT}{\mathcal T}
\newcommand{\CS}{\mathcal S}
\newcommand{\CM}{\mathcal M}

\newtheorem{innercustomthm}{Proposition}
\newenvironment{customthm}[1]
  {\renewcommand\theinnercustomthm{#1}\innercustomthm}
  {\endinnercustomthm}
\newtheorem{einnercustomthm}{Extended Theorem}
\newenvironment{ecustomthm}[1]
  {\renewcommand\theeinnercustomthm{#1}\einnercustomthm}
  {\endeinnercustomthm}

\newcommand{\red}{\textcolor{red}}
\newcommand{\blue}{\textcolor{blue}}
\newcommand{\mblue}[1]{\color{blue}{#1}\color{black}}
\makeatletter
\newcommand{\customlabel}[2]{%
\protected@write \@auxout {}{\string \newlabel {#1}{{#2}{}}}}
\makeatother


\def\qed{\hfill\vrule height4pt width4pt
depth0pt}
\def\reff #1\par{\noindent\hangindent =\parindent
\hangafter =1 #1\par}
\def\title #1{\begin{center}
{\Large {\bf #1}}
\end{center}}
\def\author #1{\begin{center} {\large #1}
\end{center}}
\def\date #1{\centerline {\large #1}}
\def\place #1{\begin{center}{\large #1}
\end{center}}

\def\date #1{\centerline {\large #1}}
\def\place #1{\begin{center}{\large #1}\end{center}}
\def\intr #1{\stackrel {\circ}{#1}}
\def\R{{\rm I\kern-1.7pt R}}
 \def\N{{\rm I}\hskip-.13em{\rm N}}
 \newcommand{\cprod}{\Pi_{i=1}^\ell}
\let\Large=\large
\let\large=\normalsize


\begin{titlepage}
\def\thefootnote{\fnsymbol{footnote}}
\vspace*{1.1in}

\title{Completeness and Transitivity of Preferences  on Mixture Sets\fn{This version of a preliminary draft circulated in January 28, 2018 was completed  during Khan's visit to the Department of Economics at the University of Queensland during July 27 to August 13, 2018.  The authors are grateful to Rabee Tourky for his insightful, off-the-cuff remarks, to Priscilla Man, Patrick O'Callaghan and John Quiggin for conversation and encouragement, and to H\"{u}lya Eraslan for her careful reading.   Khan should  also like to thank the department for its hospitality and  to acknowledge inspiration received over the years from listening to talks on this broad subject-matter by Edi Karni.}} 


\author{Tsogbadral Galaabaatar,\fn{Dept of Economics, Ryerson Univ, Toronto, ON M5B2K3.  {\bf E-mail}
{tgalaab1@economics.ryerson.ca}} 
M. Ali Khan\fn{Department of Economics, Johns Hopkins University, Baltimore, MD 21218. {\bf E-mail}
{akhan@jhu.edu}} and  
 Metin Uyan{\i}k\footnote{School of Economics, University of Queensland, Brisbane, QLD 4072.  {\bf E-mail}
{m.uyanik@uq.edu.au}}}

\vskip 1.00em
\date{September 4, 2018}

\vskip 1.75em

\vskip 1.00em

\baselineskip=.18in

\noindent{\bf Abstract:} In this paper, we show that the presence of the Archimedean and the mixture-continuity properties of a binary relation, both  empirically non-falsifiable in principle,  foreclose the possibility of consistency (transitivity) without  decisiveness  (completeness),  or decisiveness without consistency, or in the presence of a weak consistency condition, neither.   The basic result can be sharpened when specialized   from the context of a generalized mixture set to that of a mixture set in the sense of Herstein-Milnor \citeyearpar{hm53}. We relate the results to the antecedent literature, and view them as part of an investigation into  the interplay of the structure of the choice space and the behavioral assumptions on the binary relation  defined on it;  the ES research  program due to \cite{ei41} and \cite{so65}, and one to which \cite{sc71} is an especially influential contribution.    

\vskip 1.3in

\noindent {\it Journal of Economic Literature} Classification
Numbers: C00, D00, D01

\vskip .35em

\noindent {\it 2010 Mathematics Subject} Classification Numbers: 91B55, 37E05.

\vskip .35em


\vskip .35em

\noindent {\it Running Title:} Completeness and Transitivity on Mixture Sets

\end{titlepage}


\tableofcontents

\vspace{15pt}

 --------------------------------------------------------------- 

\vspace{15pt}


\setcounter{footnote}{0}


\setlength{\abovedisplayskip}{0.05cm}
\setlength{\belowdisplayskip}{0.05cm}


\newif\ifall
\alltrue 

%
%
%
%
%
%
%


\section{Introduction} \lb{sec: introduction}  

Khan-Uyan\i k \citeyearpar{ku17} highlight what they refer to as the Eilenberg-Sonnenschein (ES) research program, and present a comprehensive treatment of the two-way relationship between the properties of a binary relation and of the set over which the relation is defined.  Confining themselves exclusively to the topological register, they develop a general theory of a complete, transitive, continuous and $k$-non-trivial ($k$ any natural number)  binary relation defined on a $k$-connected topological space.  As is well-known, the transitivity  postulate in the presence of completeness  was studied  in \cite{ei41} and \cite{so65}, and completeness in the presence of transitivity by  \cite{sc71}, but all only  for $k =1.$   Khan-Uyan\i k \citeyearpar{ku17} also broaden the theory to 
include  a  sufficiency result, based  on the work of Sen (1969), that delineates   the consequences of $1$-connected (connected) choice sets for  transitivity alone of a given continuous binary relation.  All this notwithstanding, the fact remains that the bulk of modern decision theory, in of itself and in terms of applications to mathematical economics and mathematical  social sciences more generally,  does not confine itself solely to the topological register, and it is therefore natural to ask for the added consequences that can be obtained by  supplementing it  with the algebraic one.  However, this is not the question we ask here. 
 
Rather than a supplementation,  we inquire into  the reformulation of the ES program for the  setting considered  in the pioneering work of  von Neumann-Morgenstern \citeyearpar{vm47} and  \cite{ma50}, and one given a definitive treatment in Herstein-Milnor \citeyearpar{hm53}.  There,  as is well-understood,  it is not so much a question of supplementing topology  by  algebra as it is of working with a novel  mathematical setting of a mixture  space,  a setting that is endowed with a most minimal topological requirement. In particular,  instead of a topology being  assumed on the space of objects, one simply utilizes the standard Euclidean unit interval used to combine the objects.\fn{To be sure, there is a literature that  
imposes a topological structure on the objects themselves, but as we point out in Section 4 devoted to applications, this is best discussed in the setting of Khan-Uyan\i k \citeyearpar{ku17}.}  
  This  consequence of working with  a generalization of a linear to a mixture space leads, as a necessary concomitant,  to the fact that the results in Khan-Uyan\i k \citeyearpar{ku17}, assuming as they do a topology on the choice set,  have no direct application to this richer alternative set-up. New mathematical argumentation along with  its associated techniques is required.\fn{This is elaborated below in Section 5 devoted to the proofs of the results.}     Even if the choice set is taken to be the  simplex, endowed with both topological and algebraic structures, their  results cannot be directly used  and applied.  Furthermore, given that the \lq\lq topological action" involves only the unit interval, 
    the question of  working with any richer concept of connectedness and its natural extensions is excluded by 
    default.\fn{In anticipation, we may point out here that {\it pace} McCarthy-Mikkola \citeyearpar{mm18}, the results concerning mixture spaces are {\it not} purely algebraic and topologically-free. As we elaborate in the sequel, they use an algebraic condition that is  equivalent to mixture-continuity.}    And  therefore, rather than going {\it backwards} from behavioral properties of preference relations to deduce  properties of the topology on the choice set, we can only go {\it forwards} to deduce consequences for behavior of the continuity properties of the given  relation, continuity now being formalized by the Archimedean property and by mixture-continuity, both rooted in the unit interval.\fn{To explicate this a little further, it is not that the {\it backward} direction cannot be executed but that in focusing on the topologies on the unit interval, it would involve consequence of little substantive economic content.  However,  we do  go {\it backwards} in Theorem 4 and in Observation 1, not in the context of topological properties of the choice set,  but in terms of  its linear structure!}  

On moving beyond broad methodological remarks,  this paper contributes to a particular strain within the ES program.  This is the work of \cite{du11}, Karni-Safra \citeyearpar{ks15}, and especially the recent contribution of McCarthy-Mikkola \citeyearpar{mm18}.  Its trajectory can be simply laid out.  Dubra, in his specialization of  the choice set to the space of  lotteries   on a set of finite prizes, showed how mixture-continuity and the Archimedean property, together with the independence axiom, yields Schmeidler's conclusion that  a non-trivial, reflexive and transitive relation is necessarily  complete.  Karni-Safra underscore the thrust of Dubra's contribution by relaxing  independence to the \lq\lq betweenness" property or cone-monotonicity property, and again, like him,  by appealing to Schmeidler's theorem to obtain completeness of the given relation.  In a move that is not unsurprising, McCarthy-Mikkola  ignore the Karni-Safra generalization,  and revert to the original Dubra setting with the independence axiom in full operation, and generalize Dubra's result to a convex set in a linear space that is not limited to be finite-dimensional.   We apply Occam's razor, and  show that  these particular results obtain without any linearity assumption on preferences! In particular, all of the results of McCarthy-Mikkola \citeyearpar{mm18} follow as corollaries of the results reported here.\fn{Again anticipating somewhat, even though they use  an algebraic version of mixture-continuity, their condition is equivalent to the usual mixture-continuity notion under the independence assumption. Hence, their results can be restated in terms of the usual  mixture-continuity notion, and thereby generalized.  We also point out here that unlike \cite{du11} and  Karni-Safra \citeyearpar{ks15}, rather than an appeal to Schmeidler's theorem, we rely on the  {\it method of proof} of  his theorem. On all this, see the first paragraph of Section 4.  \label{fn:DKSMM}}

But this is perhaps not the primary contribution of the paper.  It is rather  to  show that once the question is set  within the broad outlines of the {\it forward} direction of the   ES program, we  obtain in the context of generalized mixture-spaces, as formulated and studied in \cite{fi82},\fn{See Fishburn's references to von Neumann-Morgenstern \citeyearpar{vm47}, \citet{fi64}, \cite{ch71} and Fishburn-Roberts \citeyearpar{fr78}.} {\it both} transitivity and completeness under a  considerably weaker version of the transitivity postulate. This articulation is then followed   up  by delineating    possibilities  opened by  a further substitution of  the weaker transitivity notion by  a convexity postulate on preferences. In terms of a more detailed overview of the results, our first theorem shows that under semi-transitivity and transitivity of the  symmetric part of a reflexive and non-trivial binary relation,    mixture-continuity and   the Archimedean property yield  both completeness and transitivity of the relation in the context of a generalized mixture space. The two properties are bundled conclusions: under the assumed hypotheses, one property cannot be obtained without the other.  Any agent cannot be consistent without being decisive, or decisive without being consistent.    Next,  we specialize the setting to a mixture space, and present a result, and its three  corollaries,  that concern only the transitivity (consistency) postulate and its various relaxations as  in \cite{se69}.\fn{Khan-Uyan\i k \citeyearpar{ku17} refer these relaxations as \lq\lq Sen's deconstruction of the transitivity postulate;" also see \cites{fi70} survey.}  Continuing with the setting of a mixture space, we bring the completeness postulate into the picture, and show that  transitivity of the symmetric part of an Archimedean or a  mixture-continuous binary relation, a  convexity assumption on preferences is  sufficient for  transitivity.   Finally, in what may be the most surprising and anti-climactic finding, we show (in Theorem 4 below) that if the preference relation is \lq\lq very nice," the model  essentially  collapses to a situation where the standard greater-than-or-equal-to relation on a unit interval is being investigated.  The consequence of this for the Herstein-Milnor representation theorem are  unmistakable, and we are thereby led directly to an alternative proof of their result.\fn{See Corollary \ref{thm: representation} below, and its proof in Section 5.}    In sum,  it is in this bundling of the results of Eilenberg, Sonnenschein, Sen and Schmeidler, Herstein-Milnor serving as an important subtext,  that each individual contribution is mutually illuminated and  allows  a maturer theory.

But mature or otherwise, the question remains as to what precisely these theorems offer in terms of the antecedent literature. How can the theory be applied?  We have already referred  to the work of Dubra, Karni, Safra, McCarthy and Mikkola (henceforth DKSMM),  but it is Section 4 below that 
we attempt a more  careful reading and systematic discussion of the  literature  with these  theorems in pure theory in hand. We do so under the criterial rubric of 
{\it redundancy} and {\it hiddenness} on the one hand, and of {\it fragility} and {\it flimsiness} on the other. 
We have already mentioned the  taking of Occam's razor to the theorems and of removing {\it redundancies} in them; the criteria of {\it hidenness} is only a little less straightforward --   the Malinvaud-Samuelson  exposing of the independence axiom in von Neumann-Morgenstern \citeyearpar{vm47} being the archetypical example.\fn{See \cite{ma52} for the example, and Khan-Uyan\i k \citeyearpar{ku17} for a more detailed explication.}   As regards the other two criteria, they are robustness criteria inspired by \cite{ge13}, and  perhaps ought to be seen as further elaboration of the  incorporated {\it hiddenness} criterion. In any case, we turn to their   formal explication and discussion below. 

\section{Notational and Conceptual Preliminaries} \lb{sec: preliminaries}   

Here, and later, lower case Greek letters will always denote real numbers in  [0,1] which  is endowed with the usual Euclidean topology. This is in keeping with the inspired usage of Herstein-Milnor \citeyearpar{hm53}.

Let $X$ be a set. A subset $\succcurlyeq$ of $X\times X$ denote a {\it binary relation} on $X.$ We denote an element $(x,y)\in ~\!\!\! \succcurlyeq$ as $x\succcurlyeq y.$ The {\it  asymmetric part} $\succ$ of $\succcurlyeq$ is defined as $x\succ y$ if $x\succcurlyeq y$ and  $y\not\succcurlyeq x$, and its {\it symmetric part} $\sim$ is defined as $x\sim y$ if $x\succcurlyeq y$ and $y\succcurlyeq x.$   We call $x\bowtie y$ if $x\not\succcurlyeq y$ and $y\not\succcurlyeq x$. The inverse of $\succcurlyeq$ is defined as  $x\preccurlyeq y$ if $y\succcurlyeq x$. Its asymmetric $\prec$ is defined analogously and its symmetric part is $\sim$.  We provide the descriptive adjectives pertaining to a relation in a tabular form for the reader's convenience in the table below. 

\begin{table}[ht]
\begin{center}
\begin{tabular}{lll} 
\hline  \noalign{\vskip 1mm}     
{\it reflexive}   &$x\succcurlyeq x$ $\forall x\in X$\\ 
{\it complete}  & $x\succcurlyeq y$ or $y\succcurlyeq x$ $\forall x,y\in X$\\ 
{\it non-trivial}   &  $\exists x,y\in X$ such that $x\succ y$\\ 
{\it transitive}  & $x\succcurlyeq y\succcurlyeq z \Rightarrow x\succcurlyeq z$ $\forall x,y,z\in X$\\ 
{\it negatively transitive} & $x\not\succcurlyeq y\not\succcurlyeq z \Rightarrow x\not\succcurlyeq z$  $\forall x,y,z\in X$ \\
  {\it semi-transitive}  & $x\succ y\sim z \Rightarrow x\succ z$ and  $x\sim y\succ z \Rightarrow x\succ z$ $\forall x,y,z\in X$ \\
\hline
\end{tabular}
\end{center}
\vspace{-14pt}

\caption{Properties of Binary Relations}
\lb{tbl: relation}
\end{table}  

Next, we provide a definition of the mixture set due to Herstein-Milnor \citeyearpar{hm53}. 
\vspace{-4pt}

\df
A set $\CS$ is said to be a {\it mixture set} if for any $x,y\in \CS$ and for any $\mu$ we can associate another element,\fn{In deference to Herstein-Milnor \citeyearpar{hm53}, lower case Greek letters consistently denote real numbers in [0,1].} which we write as $x \mu y,$ which is again in $\CS,$ and where for all $x,y\in \CS$ and all $\lambda,\mu,$ {\nf (S1)} $x1y = x$, {\nf (S2)}  $x\mu y =  y(1-\mu)x$, {\nf (S3)} $(x \mu y)\lambda  y =  x (\lambda\mu) y$.
%
%
\lb{df: mixture}
\dff

\nt Note that the following property of a mixture set is implied by S1-S3 above.\fn{See Luce-Suppes \citeyearpar[p288]{ls65} or \citet[Section 2.4]{fi82} for a proof.} 
  
\ben[topsep=3pt]
\setlength{\itemsep}{-1pt} 
\ml[{(S4)}] $(x\lambda y)\mu (x\beta y ) =  x(\mu \lambda +(1-\mu)\beta) y$ for all $x,y\in \CS$ and all $\lambda,\mu, \beta.$  
\een

\nt The notion of a   mixture set can be routinely generalized by replacing equalities between the pairs of mixtures by indifference in the definition above.\fn{This definition is due to \citet[Section 2.3]{fi82}. A complete axiomatization of a generalized mixture set is first provided, to the best of the authors' knowledge,  by \citet[p8]{fi64}. A form of (M3) is used in Luce-Raiffa \citeyearpar[p26]{lr57} in the context of the reduction of compound lotteries. Even though von Neumann-Morgenstern \citeyearpar[Section 3.6]{vm47} use equality in their algebra of combining axioms, their interpretation is consistent with the use of indifference; see also \citet{ch71} and Fishburn-Roberts \citeyearpar{fr78} for applications and discussion of generalized mixture sets.  Even though Chipman seems unaware of the work of  Sonnenschein and Schmeidler,  his discussion of the 
 Archimedean assumption with Samuelson's writings  as the relevant background, and his muted claim that continuity does not have behavioral implications, surely merits further engagement. Another reference that merits future engagement regarding applications is surely \cite{gu77}; see Footnote~\ref{fn:gu} below. \label{fn:fi}}

\df 
Let $\CM$ be a set and $\succcurlyeq$ a reflexive binary relation on it with a transitive symmetric part $\sim.$ Then, $\CM$ is said to be a {\it generalized mixture set} {\nf (}induced by $\sim${\nf )} if for any $x,y\in \CM$ and for any $\mu$ we can associate another element, which we write as $x \mu y,$ which is again in $\CM,$ and where for all $x,y\in \CM$ and all $\lambda,\mu, \beta,$  {\nf (M1)} $x1y \sim  x,$ {\nf (M2)} $x\mu y \sim   y(1-\mu)x,$ {\nf (M3)} $(x \mu y)\lambda  y \sim  x (\lambda\mu) y,$  
{\nf (M4)} $(x\lambda y)\mu (x\beta y ) \sim  x(\mu \lambda +(1-\mu)\beta) y$. 
\lb{df: gmixture}
\dff 

Next, we turn to the various properties of a binary relation on a generalized mixture set, and develop the following notation for subsets of $[0,1].$ For any  $\succcurlyeq$ on  $\CM$ and for any $x,y,z\in\CM,$ let   
$$
\begin{array}{c}
A_\succcurlyeq(x,y,z)=  \{\lambda~|~x\lambda y \succcurlyeq z\} ~\text{ and }~ A_\preccurlyeq(x,y,z)= \{\lambda~|~z\succcurlyeq x\lambda y \}.
\end{array}
$$
The sets  $A_\succ(x,y,z),  ~\! A_\prec(x,y,z), ~\!  A_\sim(x,y,z)$ and $A_{\bowtie}(x,y,z)$ are analogously defined.

\df
We call a binary relation $\succcurlyeq$ on a generalized mixture set $\CM$ 
\ben[{\nf (i)}, topsep=0pt]
\setlength{\itemsep}{-3pt} 
\ml mixture-continuous if for all $x,y,z\in \CM,$ the sets $A_\succcurlyeq(x,y,z)$ and $A_\preccurlyeq(x,y,z)$ are closed; 

\ml Archimedean if for all $x,y,z,w\in \CM$ with $x\succ y$, $x\bowtie w$ and $y\bowtie z$ there exist $\lambda, \delta \in (0,1)$  such that $\lambda\in A_\succ(x,z,y)$ and   $\delta\in A_\prec(y,w,x);$ 

\ml strongly Archimedean if for all $x,y,z\in \CM$ with $x\succ y$, there exists $\lambda, \delta \in (0,1)$  such that $\lambda\in A_\succ(x,z,y)$ and   $\delta\in A_\prec(y,z,x)$.
\een
\lb{df: properties}
\dff

\nt Note that the Archimedean property above is weaker than strong Archimedean property -- the latter is the version that is usually assumed in the literature. 


\df
A binary relation $\succcurlyeq$ on a generalized mixture set $\CM$ is 
\ben[{\nf (i)}, topsep=3pt]
\setlength{\itemsep}{-1pt} 

\ml linear if for all $x,y\in \CM,$  all $z\in \{x,y\}$ and all $\lambda\in (0,1]$  $~x\sim y$ if and only if $\lambda\in A_\sim(x,z,y\lambda z),$

\ml convex if for all $x,y,z\in \CM$  and all $\lambda,$  $x\succcurlyeq z$ and $y\succcurlyeq z$ implies $\lambda\in A_\succcurlyeq (x,y,z),$ 

\ml concave if for all $x,y,z\in \CM$  and all $\lambda,$  $z\succcurlyeq x$ and $z\succcurlyeq y$ implies $\lambda\in A_\preccurlyeq (x,y,z),$ 

\ml star-convex if for all distinct $x, y\in \CM$  and all $\lambda\in (0,1),$  if  $x\succcurlyeq y,$ then $\lambda\in A_\succ (x,y,y),$ 

\ml star-concave if for all distinct $x,y\in \CM$  and all $\lambda\in (0,1),$  if  $y\succcurlyeq x,$ then $\lambda\in A_\prec (x,y,y).$ 
\een
\lb{df: properties2}
\dff
\vspace{-5pt}


It is well-known that the conventional  independence assumption, or the weaker property of {\it betweenness,}   implies that the preference relation is linear.  
  Under completeness and transitivity, 
 star-convexity implies convexity and star-concavity implies concavity. 
  However, without the completeness assumption, there is no inclusion relationship between convexity and star-convexity as well as between concavity and star-concavity. 
   For example, any preference relation with thick indifference curves illustrates that convexity does not imply star-convexity.  
 In order to see that star-convexity does not imply convexity, let $X=\{x\in \Re^3_+| \sum_i x_i=1\}.$ Assume $\succcurlyeq$ is a reflexive binary relation on $X$ such that $(1,0,0)\lambda (0,1,0)\succ (0,1,0)$ and $(0,0,1)\lambda (0,1,0)\succ (0,1,0)$ for all $\lambda\in (0,1]$. There are no other comparable points. Then, it is clear that $\succcurlyeq$ is star-convex and not convex. Analogous arguments illustrate that there is no inclusion relationship between concavity and star-concavity. 


Our laying-out of the conceptual preliminaries would not be complete without any mention of results that explore the somewhat more subtle converse to the assertion that every convex set is a mixture set. The fact that the converse  does not hold is hardly esoteric, and deserves to be more widely  known in the mathematical social science literature; see for example \citet[Section V.II]{wa89} and \citet[p61]{mo01}. \citet[Theorem 2]{st49} and \citet[Theorems 3.2 and 3.4]{ha54} provide axiomatic characterization of convex sets by  showing that a mixture set $\CS$ is isomorphic to a convex subset of some linear space if and only if it satisfies the following two axioms.\fn{See \citet{gu77} for an expository article on abstract convexity and its applications to behavioral, social and physical sciences.    Note that there is no reference to Stone in Hausner's paper, and given that Mongin is also silent about the relationship between the two works, we regard them as independent.  \label{fn:gu}}

\ben[{\nf (i)}, topsep=3pt]
\setlength{\itemsep}{1pt} 
\ml[(C1)]  {\it $\!\!\text{ For all } x\in \CS \text{ and all }  \lambda \in (0,1), x\lambda y =x\lambda y' \text{ implies } y=y'$.} \lb{it: c1}
\ml[(C2)] {\it $\!\!\text{ For all } x,y,z\in \CS \text{ and all } \lambda,\mu\in [0,1] \text{ with } \lambda\mu\neq 1, (x\lambda y) \mu z= x (\lambda \mu) \left(y \displaystyle \frac{\mu(1-\lambda)}{1-\lambda\mu}z\right)$.} \lb{it: c2}
\een


\nt In a more recent work,  \citet{mo01} introduces the concept of non-degeneracy and shows that it is equivalent to (C1)-(C2). A function $u:\CS\ra \Re$ is said to be {\it mixture preserving}  if $
\text{for all } x,y\in \CS \text{ and all } \lambda \in [0,1], u(x\lambda y)=\lambda u(x) + (1-\lambda) u(y).
$
Denote by $\mathcal L(\CS)$ the set of  all mixture preserving functions on $\CS,$ and define a relation $\approx$ on $\CS$ as 
$
x\approx y \text{ if and only if for all } u\in \mathcal L(\CS), u(x)=u(y). 
$
It is clear that $\mathcal L(\CS)$ is a vector space,  $\approx$ is an equivalence relation, and the quotient space $\CS|\!\!\approx$ is a mixture set with the mixture operation $[x]\lambda [y]=[x\lambda y]$ for all $[x], [y]\in \CS|\!\approx$ and all $\lambda\in [0,1]$.

\df
A mixture set $\CS$ is non-degenerate if all classes of $\CS|\!\approx$ are singletons, i.e.,  for all $x,y\in \CS$, $x\approx y$ implies $x=y$. 
\lb{thm: non-degenerate}
\dff

\nt The following proposition encapsulates  the main results of Stone-Hausner-Mongin and is copied from  \citet[Proposition]{mo01} for the reader's convenience.

\begin{customthm}{0}[Stone-Hausner-Mongin]
The following are equivalent for any  mixture set $\CS$.  
\ben[{\nf (a)},  topsep=3pt]
\setlength{\itemsep}{-1pt} 
\ml $\CS$ satisfies (C1) and (C2). \lb{it: mc1}
\ml $\CS$ is non-degenerate. \lb{it: mc2}
\ml $\CS$ is isomorphic to a convex subset of some linear space.  \lb{it: mc3}
\een
\lb{thm: shm}     \label{prp: shm}
\end{customthm}


We end this section by presenting the relationship between  Archimedean and mixture-continuity properties without the completeness and full transitivity assumptions.   
 We first show that  Archimedean property is equivalent to a topological condition under mixture-continuity. 

\prp
  Let $\succcurlyeq$ be a semi-transitive binary relation on a generalized mixture set $\CM$ with mixture-continuity. Then, the following are equivalent. 
  \ben[{\nf (a)}, topsep=1pt]
\setlength{\itemsep}{-3pt} 
\ml $\succcurlyeq$ is Archimedean, \lb{it: arch}
\ml $\succcurlyeq$ is strongly Archimedean, \lb{it: sarch}
\ml $A_\succ(x,y,z)$ and $A_\prec(x,y,z)$ are open for all $x,y,z\in \CM.$ \lb{it: open}
\een 
\lb{thm: arc}
\prpp
\vspace{-3pt}

\noindent When preferences are complete,  mixture-continuity and condition \ref{it: open} above are equivalent, and hence the Archimedean properties follow from mixture-continuity without any further assumptions.  However, without the completeness assumption, mixture-continuity and condition \ref{it: open} are independent. It is clear from the proof that, assertion \ref{it: open}$\Rightarrow$\ref{it: arch}  does not require the mixture-continuity assumption, hence condition \ref{it: open} is stronger than the Archimedean properties.\fn{Moreover, in mixture sets, this direction of the equivalence result  does not require semi-transitivity; indeed,  assertion \ref{it: arch}$\Rightarrow$\ref{it: open} does not require semi-transitivity even in generalized mixture sets.} We elaborate these points in the Appendix by providing examples. 
\smallskip

  
The  Archimedean property is weaker  than mixture-continuity,  even under the completeness and transitivity assumptions being in force -- either a convexity condition, or a further continuity property,  needs to be  imposed on preferences in order to obtain 
 mixture-continuity of an Archimedean relation.\fn{See for example \citet{du11} and Karni-Safra \citeyearpar{ks15} for the former, and  \citet{ka07} for the latter.} The following result provides sufficient conditions for mixture-continuity under the strong Archimedean property without completeness, full transitivity and convexity of preferences.

\prp
Let $\succcurlyeq$ be a semi-transitive and strongly Archimedean binary relation on a generalized mixture set $\CM$ such that for all $x,y,z\in \CM$, $A_{\bowtie}(x,y,z)$ is open, and $A_\succcurlyeq(x,y,z)$ and $A_\preccurlyeq(x,y,z)$ have finitely many components. Then, $\succcurlyeq$ is mixture-continuous. 
\lb{thm: mc}
\prpp

\nt Under the completeness assumption, openness of $A_{\bowtie}(x,y,z)$ trivially holds.\fn{As in Proposition \ref{thm: arc}, semi-transitivity is not needed for this proposition in mixture sets.} The finiteness of the components is implied by concavity and convexity of the preference relation, which are satisfied under independence hypothesis.  We show in the Appendix that each of the assumptions of this proposition is not redundant.  

\vspace{-7pt}

\section{The Results: Presentation and Discussion}\lb{sec: main}

In this section, we present four theorems and with the help of four observations draw out four  corollaries from them. We begin with  our first result that pertains to a  generalized mixture set. 

\thm
Any non-trivial, reflexive, semi-transitive, mixture-continuous and Archimedean binary relation $\succcurlyeq$ with a transitive symmetric part $\sim$ on a generalized mixture set induced by $\sim$, is complete and transitive.
\lb{thm: ct}
\thmm
 
\nt For the case of an anti-symmetric relation, we can re-state the above result  without any reference  to any form of transitivity.  

\obs Any anti-symmetric, non-trivial, reflexive, mixture-continuous and Archimedean binary relation $\succcurlyeq$ on a generalized mixture set induced by $\sim$, is complete and transitive.
\lb{obs: anti-symmetric}
\obss

\nt A simple elaboration shows that any anti-symmetric relation satisfies semi-transitivity and its symmetric part is transitive.    The reader who does not want to worry about  different transitivity concepts\fn{Khan-Uyan\i k \citeyearpar[Section 3]{ku17} discuss in detail the relationship between different transitivity concepts.}  can refer to this simpler version of the theorem. Moreover, this observation has important implications  on the structure of the mixture set to which we return at the end of this section. The following corollary shows that, under completeness assumption, a weak form of transitivity implies full transitivity. 

\cor
Any complete, mixture-continuous and Archimedean binary relation $\succcurlyeq$ on a generalized mixture set $\CM$  induced by $\sim$, is transitive if any or both of the following holds: 
\ben[{\nf (a)}, topsep=1pt]
\setlength{\itemsep}{-3pt} 
\ml $x\succ y\sim z$ implies $x\succ z$ for all $x,y,z\in \CM$,  \lb{it: rsemi}
\ml $x\sim y\succ z$ implies $x\succ z$ for all $x,y,z\in \CM$.  \lb{it: lsemi}
\een
\lb{thm: mixturest}
\corr

\vspace{-4pt}

Our second  result shows that the transitivity of the asymmetric part of a  complete, strongly Archimedean and star-convex (or star-concave) binary relation is sufficient for its transitivity. 
\vspace{-4pt}

\thm
Any complete and strongly Archimedean binary relation $\succcurlyeq$ with a transitive asymmetric part $\sim$ on a generalized mixture set induced by $\sim$, is transitive if any or both of the following holds: 
\ben[{\nf (a)}, topsep=1pt]
\setlength{\itemsep}{-3pt} 
\ml $\succcurlyeq$ is star-convex,   \lb{it: sconvex}
\ml $\succcurlyeq$ is star-concave.    \lb{it: sconcave}
\een
\lb{thm: mixturecat}
\thmm

 \vspace{-4pt}
 
Our third result  shows that, in a mixture set, 
certain convexity properties are sufficient for semi-transitivity.
\vspace{-5pt}

\thm The following are true for a reflexive, mixture-continuous and Archimedean binary relation $\succcurlyeq$ on a mixture set $\CS$ whose symmetric part is transitive.  
\ben[{\nf (a)}, topsep=1pt]
\setlength{\itemsep}{-3pt} 
\ml If $~\!\! \succcurlyeq$ is linear, then  it is semi-transitive. \lb{it: linear}
\ml If $~\!\! \succcurlyeq$ is convex, then $x\sim y\succ z$ implies $x\succ z$ for all $x,y,z\in \CS$. \lb{it: convex}
\ml If $~\!\! \succcurlyeq$ is concave, then $x\succ y\sim z$ implies $x\succ z$ for all $x,y,z\in \CS$. \lb{it: concave}
\ml If $~\!\! \succcurlyeq$ is complete, then its convexity or its concavity implies its semi-transitivity. \lb{it: complete}
\een
\lb{thm: comparison}
\thmm

\vspace{-4pt}

 We now rely on the following observation to bring out the fuller implications of the result. 
\vspace{-20pt}

\obs If a relation is convex and concave, then it is linear. The converse is true under mixture continuity and the Archimedean axiom.\fn{See Lemma 1 in the Appendix below for a proof.   \lb{obs: convex}}
\obss
\vspace{-4pt}

\nt Figure \ref{fig: comparison} below illustrates examples of linear, convex, concave and semi-transitive preferences. It is easy to see that, when the choice set is a convex subset of a linear space, linearity implies that the indifference sets are convex (thick indifference curves are allowed), convexity that the upper section of the weak preference relation is convex and concavity that the lower section of the weak preference relation is convex. 

\begin{figure}[ht]
\begin{center}
  \includegraphics[width=6.3in, height=1.15in]{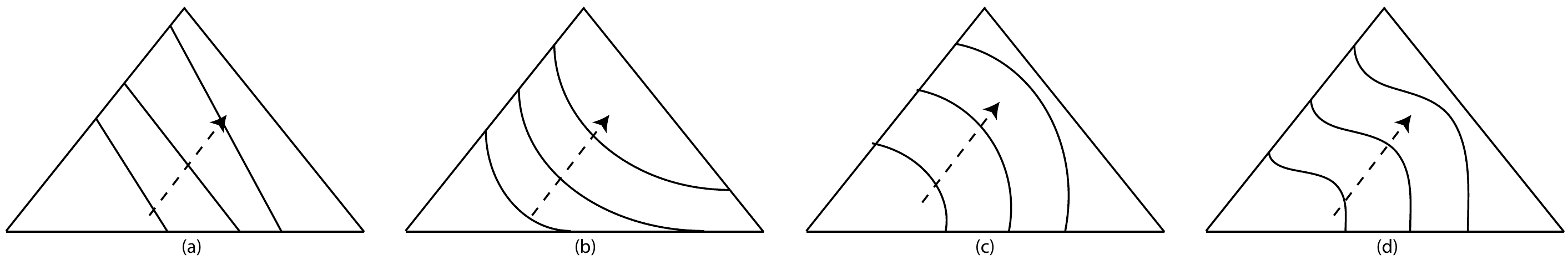}
    \end{center}   
    \vspace{-10pt}
    
   \footnotesize{$X=\{x\in \Re_+^3|~\sum_{x_i}=1\}$ is the two dimensional unit simplex. The curves illustrate the indifference curves and the arrow indicates the direction of preferences. Panel (a) illustrates a linear preference relation, (b) convex but neither linear, nor concave (c) concave but neither linear, nor convex and (d) semi-transitive that is not linear, convex and concave. } 
\vspace{-5pt}   

\caption{Linear, Convex, Concave and Semi-transitive Preferences}
\vspace{-5pt}   
   
  \lb{fig: comparison}
\end{figure}


 The following two corollaries  illustrate that, in a mixture set, semi-transitivity can be substituted by a convexity property in the hypothesis of Theorem \ref{thm: ct}. First, we show that we can replace semi-transitivity with linearity. 
\vspace{-5pt}   

\cor
Any non-trivial, reflexive, mixture-continuous and Archimedean binary relation $\succcurlyeq$ with a transitive symmetric part on a mixture set, is complete and transitive if any or both  of the following holds: 
\ben[{\nf (a)}, topsep=1pt]
\setlength{\itemsep}{-3pt} 
\ml  $\succcurlyeq$ is linear, 
\ml $\succcurlyeq$ is semi-transitive. 
\een
\lb{thm: mixture}
\corr
\vspace{-3pt}   


\nt Next, we show that linearity hypothesis in the corollary above can be replaced with one of convexity or concavity in the presence of completeness. 

\cor
Any complete, mixture-continuous and Archimedean binary relation $\succcurlyeq$ with a transitive symmetric part  on a mixture set, is transitive if any or both of the following holds: 
\ben[{\nf (a)}, topsep=1pt]
\setlength{\itemsep}{-3pt} 
\ml  $\succcurlyeq$ is convex, 
\ml $\succcurlyeq$ is concave. 
\een 
\lb{thm: mixturecc}
\corr


We end this section by presenting a joint implication of Theorem \ref{thm: ct} and Observation \ref{obs: anti-symmetric}.

\thm 
If there exists a non-trivial, complete, transitive, anti-symmetric and mixture continuous binary relation $\succcurlyeq$ on a generalized mixture set $\CM$ induced by $\sim$, then $\CM$ is isomorphic to an interval in $\Re$ and $\succcurlyeq$ is equivalent to the usual ``greater-than-or-equal-to'' or ``less-than-or-equal-to'' relation.
\lb{thm: realline}
\thmm
\vspace{-4pt}
 
\nt Theorem \ref{thm: realline} implies that $(R,\geq(\leq))$ is the only linear space with binary relation(s) that satisfies all of the above properties, hence  it provides a characterization result for $(R,\geq(\leq))$. It follows from Theorem \ref{thm: ct} that we can equivalently state the theorem above as follows.\fn{It follows from the definition of an anti-symmetric relation that if a generalized mixture set $\CM$ is induced by an anti-symmetric relation, then $\CM$ is a mixture set. Hence, we can state Theorem \ref{thm: realline} with a mixture set without loss of generality. Moreover, the non-triviality assumption in Theorem \ref{thm: realline} is not restrictive. If a binary relation satisfying other assumptions of the theorem is trivial, then the space contains at most one element, and is hence isomorphic to a (possibly empty) interval in $\Re$.}

\obs 
If there exists a  non-trivial, reflexive, anti-symmetric, mixture continuous and Archimedean binary relation $\succcurlyeq$ on a generalized mixture set $\CM$ induced by $\sim$, then $\CM$ is isomorphic to an interval in $\Re$  and $\succcurlyeq$ is equivalent to the usual ``greater-than-or-equal-to" or ``less-than-or-equal-to" relation.
\lb{obs: realline}
\obss

 A natural question arises at this stage. Is it possible to obtain a version of the result above by dropping the anti-symmetry assumption?   The example below answers this question in the negative by illustrating a complete, transitive and mixture continuous preference relation on a mixture set which is not isomorphic to a convex set in some linear space, and yet its quotient space is isomorphic to an interval in $\Re$. Moreover, the preference relation satisfies the independence axiom, hence it satisfies all assumptions of Herstein-Milnor. As already emphasized in the introduction, this is an important result, one that provides a litmus-test for evaluating results that are set up in what appears to be a generalized setting.  

\begin{figure}[ht]
\begin{center}
  \includegraphics[width=1.8in, height=1.3in]{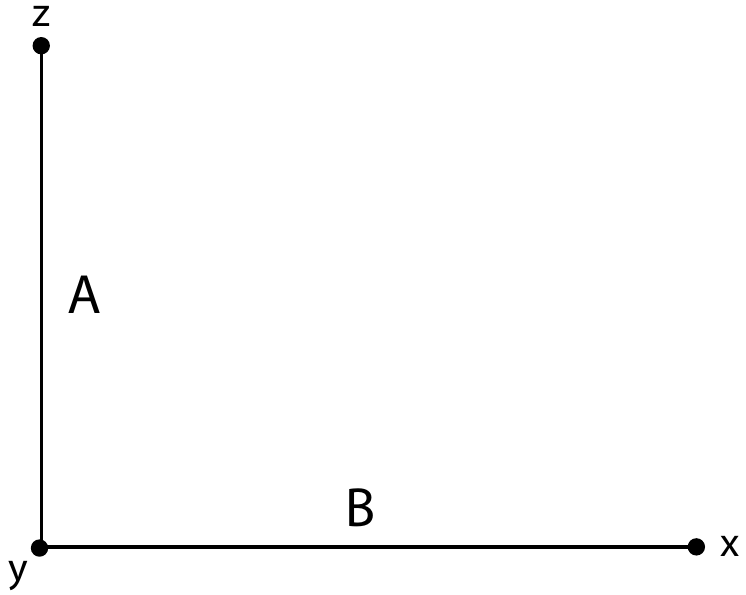}
    \end{center}   
    \vspace{-15pt}
    

\caption{A Non-convex Mixture Set with a Relation satisfying HM Axioms}
  
  \lb{fig: quotient}
\end{figure}
\vspace{-5pt}   

\exm {\nf  Let $A=\{(x_1,x_2)\in \Re^2|x_1=0, 0\leq x_2\leq 1\}$, $B=\{(x_1,x_2)\in \Re^2|0<x_1\leq 1, x_2=0\}$ and $\CS=A\cup B$. Set $x=(1,0), y=(0,0), z=(0,1)$. Define the mixture operation as follows: if $a,a'\in A$ or $a,a'\in B$, then $a\lambda a'=\lambda a + (1-\lambda)a'$ for all $\lambda\in [0,1]$; if $a\in A$ and $b\in B$, then $a\lambda b=\lambda y + (1-\lambda)b$ for all $\lambda\in [0,1)$, $a1b=a$ and $b\lambda a=a(1-\lambda) b$ for all $\lambda\in [0,1]$. Then, $\CS$ is a mixture set. Define a binary relation $\succcurlyeq$ on $\CS$ as follows: any pair in $A$ are indifferent, any point in $A$ is strictly worse than any point in $B$ and $x\lambda y \succcurlyeq x\delta y$ if and only if $\lambda\geq\delta$. It is easy to see that $\succcurlyeq$ is non-trivial, complete, transitive, mixture continuous and satisfies independence axiom It follows from $x\lambda z=x\lambda y$ for all $\lambda\in (0,1)$ and $z\neq y$ that axiom (C1) does not hold. Therefore, Proposition \ref{thm: shm} implies $\CS$ is not isomorphic to a convex set.
}\exmm

We have not yet formally defined the well-known notion of an {\it independent} relation: for all $x,y,z\in \CS$ and all $\lambda \in (0,1]$, $x\sim y$ if and only if $x\lambda z\sim y\lambda z$.   We can now derive the classic expected utility representation theorem of Herstein-Milnor on a generalized mixture set as a consequence of Theorem 4. 

\cor
Let $\succcurlyeq$ be a complete, transitive and mixture continuous binary relation  with the independence axiom on a generalized mixture set $\CM$ induced by $\sim$. Then, there exists a function $u:\CM \ra \Re$ such that for all $x,y\in \CM$, 
$$
x\succcurlyeq y \Longleftrightarrow u(x)\geq u(y)
$$ 
where $u(x\lambda y)=\lambda u(x) + (1-\lambda)u(y)$  for all $\lambda\in [0,1]$.
\lb{thm: representation} 
\corr

\nt The alternative proof is relegated to Section 5, and here we limit ourselves to the observation that is responsible for the basic idea underlying the alternative proof.

\obs In the context of the objects in Corollary~\ref{thm: representation},  it is easy to show that the quotient space $\CM|\!\sim$ is a mixture set with the mixture operation defined as  $[x]\lambda [y]=[x\lambda y]$ for all $[x], [y]\in \CM|\!\sim$ and all $\lambda\in [0,1]$.  Moreover, the derived relation $\hat \succcurlyeq$ on the quotient set,  defined as $[x]\hat \succcurlyeq [y]$ if $x'\succcurlyeq y'$ for all $x'\sim x, y'\sim y$,  is anti-symmetric, complete, transitive and mixture continuous.   
 \lb{obs: representation}
\obss
\vspace{-15pt}

\nt It is worth noting that our alternative proof is not straightforward, an is directly based on the embedding theorem of Stone-Hausner-Mongin presented as Proposition~\ref{thm: shm}. This being said, the proof itself is not difficult and bears comparison with the proofs presented in Sections 2.4 and 2.5 in \cite{fi82}.\fn{\citet[p. 20]{fi82} writes, \lq\lq  The proof of Theorem 2 [representation in generalized mixture sets] is similar to the proof given above for the construction of linear, order-preserving utilities on the basis of M1-M3 and J1-J5. Our  main  concern  in modifying the preceding proof is to make sure that the uses of $=$ from M1-M3 can be replaced by $\sim$ on the basis of the axioms in Theorem 2. [T]he construction of the desired $u$ then parallels the construction given  above with a few changes from $=$ to $\sim$, and the uniqueness proof is likewise straightforward." In our lterantive proof, we bypass the construction, however natural, and all the checking that it requires.  Also see Footnote~\ref{fn:fi}  in the context of this textual exegesis.}


\section{Implications for  the Antecedent Literature} \lb{sec: app}

We begin this section with the results of Dubra, Karni-Safra and McCarthy-Mikkola (DKSMM),  already referred to in the introduction. Theorem \ref{thm: ct} considerably generalizes these results by dropping any form of convexity assumption on preferences, by weakening transitivity and by allowing the choice space to be a generalized mixture set. Moreover, Corollary \ref{thm: mixture} shows that when the choice space is a mixture set, the semitransitivity of the preference relation  can be substituted by its linearity, which is implied  both by the {\it independence} and {\it betweenness} properties assumed in the papers above.  To elaborate a little more,  Dubra uses the independence assumption to show that preferences satisfy Schmeidler's continuity assumption, and thereby deduces his result as a corollary of Schmeidler's theorem;  KS  show that {\it independence} can be replaced by {\it betweenness} or {\it 
cone-monotonicity,} with the same method of proof.  Dubra's argumentation is  based on  \citet[Theorem 6.1]{ro70}, a result  Rockafellar refers to  as ``fundamental". 
 MM generalize Dubra's theorem to convex subsets of arbitrary real linear spaces. They use equivalent algebraic versions of the continuity assumptions and use an algebraic proof technique, and they do not use Schmeidler's theorem for their argument.  

Next, we provide some application of our results to the antecedent literature which highlights the {\it hiddenness} of  completeness and transitivity.   von Neumann-Morgenstern \citeyearpar{vm47}, Herstein-Milnor \citeyearpar{hm53} and their followers show that the following four conditions are necessary and sufficient for representation of preferences with an expected utility function: completeness, transitivity, independence, mixture-continuity and Archimedean; see \citet{fi70, fi82}, \citet{kr88} and \citet{gi09} for a survey. Our results show that a weak form of the transitivity postulate,  along with the two continuity properties,  implies both completeness {\it and} transitivity; hence both of them are hidden assumptions.  It is also worth pointing out in this connection that even though each paper in this line of literature assumes either one of these two continuity assumptions, it is by now well known that they are equivalent under the completeness and independence hypotheses. Furthermore, there  is one other delicate point worth stressing:  this is that the HM theorem is, as  stated,  false without   completeness even if we keep transitivity. Nevertheless, in the presence of the other HM assumptions, we know that mixture-continuity implies the Archimedean property, and hence adding this property  into the statement of the theorem is non-restrictive, i.e., the hypotheses of the two theorems are equivalent. As such, in this version of theorem, completeness and full transitivity are  hidden.  An analogous observation applies  to \citet{aa63}, and we leave it to the reader to reflect more generally on its implication for non-expected utility representations, as in Machina and his followers.

 In our applications, we show that the models with incomplete preferences either lack mixture-continuity or the Archimedean property.  \citet[p. 453]{au62} finds that either are  ``equally plausible, and there is no reason to prefer one over the other", and  in particular, adds:  
 
 \bqu  I personally believe the archimidean ({\it sic}) principle to be very compelling, not withstanding some of the counter-intuitive examples that have been offered in the literature. [T]here may certainly be situations in which the lexicographic order or something similar constitutes the most convenient model, so it is desirable to have a theory that covers it.\fn{See \citet[Footnote 25]{au62}.  We invite the reader 
 to use Propositions \ref{thm: arc},  \ref{thm: mc} and Footnote \ref{fn:DKSMM} above  
   to see the relationship between the continuity assumptions of this paper and those in  \citet[(4.1) and (4.2)]{au62}.}   \equ 
 
 \nt   There is surely no problem with this given the rich and considerable analysis of incomplete preferences that is now available in the literature.  However, the issue from the point of view being emphasized in this paper is slightly deeper than this.   It is not a matter of the literature lacking   one of the two assumptions but also of {\it violating,} by necessity,  one of the two equally plausible alternatives they represent. If not,  our results show that the preferences are necessarily complete!  Thus, we tend to see the results that we present above  in a somewhat   ``negative" vein  
in the modeling of incomplete and/or non-transitive preferences:    full continuity does not  allow incompleteness and/or non-transitivity of the preferences. In any case, the seemingly rather innocuous continuity assumptions have both behavioral and empirical implications.  It is presumably for this reason of strong continuity assumptions precluding the  study of such questions that   models with incomplete preferences in decision theory do not assume full continuity, but  only  one of the two continuity 
axioms.\fn{See \citet{ka14} for a recent survey and Hara et al. \citeyearpar{hor15} for the
 state-of-the-art results in this line of literature.}  This sacrifice of continuity imposes an undesirable property on preferences regarding which we  attempt a conceptual extraction.  We  study the robustness of the structure of such preferences by introducing the concepts of {\it fragility} and {\it flimsiness}:  we identify those that  violate the Archimedean property as {\it fragile}, and those which violates mixture-continuity as {\it flimsy}, and take each in turn.


\df
A binary relation $\succcurlyeq$ on a generalized mixture set $\CM$ is fragile if there exist $x,y,z\in \CM$ and  $\lambda \in A_\succ(x,y,z)\cup A_\prec(x,y,z)$ such that every open neighborhood of $\lambda$ contains a non-empty open set $V$ such that $V \subset  A_{\bowtie}(x,y,z).$
\lb{df: fragile}
\dff

The concept of fragility is first introduced by \citet{ge13} in the context of a topological space. He showed that dropping one of the continuity assumption of Schmeidler's theorem yields an undesirable case of incompleteness. Our fragility concept is motivated by Gerasimou's work. The following simple example illustrates a fragile relation.

\exm {\nf 
Assume a decision maker chooses between two alternatives, $a$ and $b$. Let $X=[0,1]$ be the set of all probability distributions on $\{a,b\}$ where $x\in X$ denote the probability of $a$ and $(1-x)$ is the probability of $b$. Assume the agent has a reflexive preference relation $\preccurlyeq$ on $X$ such that $0\prec 1$, i.e., she strictly prefers $a$ for sure to $b$ for sure. No other alternatives are comparable.  Note that every neighborhood of 1 contains an open  interval of alternatives which are incomparable to 0 and vice-a-versa.  Therefore, even though she prefers $a$ to $b$, she cannot compare $a$ with any lottery that assigns a slightly positive weight on $a$. In this example, it is clear that all assumptions of  Theorem \ref{thm: ct} hold except the Archimedean property. However, if we add Archimedean, then we know that the preference relation has to be complete. 
}\lb{ex: fragile}
\exmm

\nt The following proposition shows that dropping Archimedean assumption from Theorem \ref{thm: ct} yields a fragile preference relation. 

\prp
Any incomplete, non-trivial, reflexive, transitive and mixture-continuous binary relation on a generalized mixture set $\CM$, is fragile. 
\lb{thm: fragile} 
\prpp


\rmk {\rm  We leave it to the reader to check that  Shapley-Baucells \citeyearpar{sb98}  drop the Archimedean property and assume mixture continuity, and hence, that  the preferences in their results are fragile.  Indeed, there is an extensive literature which drops the Archimedean postulate and assumes a continuity assumption that is stronger than mixture-continuity; see for example  Ghirardato et al. \citeyearpar{gmms03}, Dubra et al. \citeyearpar{dmo04}, Evren-Ok \citeyearpar{eo11}, Ok et al. \citeyearpar{oor12}. In all these papers, the choice set is endowed with a topological structure and the sections of the weak  preference relation is closed.\fn{For a comprehensive discussion of the  structure of incomplete preferences on a topological space without any algebraic structure; see for example  Khan-Uyan\i k \citeyearpar{ku17}.}} \rmkk

For the models which drop mixture-continuity but keep strong Archimedean property, the following concept is useful, which is originally due to Khan-Uyan\i k \citeyearpar{ku17}. 

\df
A binary relation $\succcurlyeq$ on a generalized mixture set $\CM$ is flimsy if there exist $x,y,z\in \CM$ and  $\lambda \in A_{\bowtie}(x,y,z)$ such that  
  every open neighborhood $V_\lambda$ of $\lambda$ contains $\lambda'$ in $A_\succcurlyeq(x,y,z)\cup A_\preccurlyeq(x,y,z)$.
\lb{df: flimsy}
\dff

\nt Flimsiness implies that limit of some  sequence of comparable alternatives is non-comparable.  The following simple example illustrates a flimsy preference relation. 

\exm{\nf 
 Let $X=[0,3]$ and the agent has a reflexive preference relation on $X$ which satisfy the following: any pair in $[0,1)$ is indifferent to each other; similarly any pair in $(2,3]$ is indifferent to each other; lastly any point in the first set is strictly worse than any point in the latter.  No other points are comparable. Alternative 1 is non-comparable to any alternative above it but any neighborhood of it contains comparable points. Hence, it is flimsy.  In this example, it is clear that all assumptions of Theorem 1 hold except mixture-continuity. 
}\lb{ex: flimsy}
\exmm

\prp 
Any incomplete, non-trivial, reflexive, transitive, strongly Archimedean binary relation $\succcurlyeq$ on a generalized mixture set induced by $\sim$ such that $A_\succcurlyeq(x,y,z)$ and $A_\preccurlyeq(x,y,z)$ have finitely many components for all $x,y,z$, is flimsy.
\lb{thm: flimsy}
\prpp


\rmk  {\rm  Note that papers that drop mixture-continuity and assume the Archimedean property typically assume the strict preference relation to be the relevant  primitive, and  hence, strictly speaking,  our flimsiness result does not apply; see for example \citet{be02}, Manzini-Mariotti \citeyearpar{mm08}, Galaabaatar-Karni \citeyearpar{gk12, gk13} and  \citet{ev14}. The earlier works of  \citet{au62} and  \citet{ka63} do assume the weak preference relation as primitive, but their continuity assumption is weaker than the Archimedean and mixture-continuity versions we assume. Nevertheless, it is worth emphasizing that the conceptual notions of  fragility and flimsiness can be reworked and re-calibrated to apply to them.} \rmkk 


\setlength{\abovedisplayskip}{5pt}
\setlength{\belowdisplayskip}{5pt}

\section{Proofs of the Results} \lb{sec: proofs}   


We provide the proofs in the same  order in which   the results presented above in Sections 3 and 4 above.

\prf[Proof of Proposition \ref{thm: arc}]  
\noindent $\mathbf{\ref{it: arch}\Rightarrow \ref{it: open}}$ Assume $\succcurlyeq$ is mixture-continuous and Archimedean. Pick $x,y,z\in \CM.$ If $A_\succ(x,y,z)$ is empty, then it is open. Otherwise, pick $\lambda\in A_\succ(x,y,z).$ It follows from mixture-continuity and  $\lambda \notin A_\preccurlyeq(x,y,z)$ that there exists $t>0$ such that $N_t(\lambda)=\{\beta~|~|\beta-\lambda |<t\}$ is contained in the complement of $A_\preccurlyeq(x,y,z).$

Assume  there exists $\beta\in N_t(\lambda)\cap A_{\bowtie}(x,y,z).$ It follows from mixture-continuity that $A_{\bowtie}(x,y,z)$ is open. Therefore, as an open set in [0,1], $A_{\bowtie}(x,y,z)$ is union of at most countably many mutually disjoint open intervals such that the intervals of type $[0,\alpha)$ or  $(\alpha, 1]$ are allowed for any $\alpha\in (0,1)$.  
 Among these open intervals, there exists open interval $I$ such that $\beta\in I.$ If $\beta>\lambda,$ then set $\delta=\text{inf } I,$ otherwise $\delta=\text{sup } I.$ Then $\delta\in A_\succ(x,y,z).$ It follows from M4 that $(x\delta y)\gamma (x\beta y)\sim x(\gamma \delta + (1-\gamma)\beta)y$ for all $\gamma.$ Since $(\delta, \beta)\subset I,$ therefore $\gamma \delta + (1-\gamma)\beta\in A_{\bowtie}(x,y,z)$ for all $\gamma\in (0,1).$ This furnishes us a contradiction with $\succcurlyeq$ is Archimedean.  Therefore, $A_\succ(x,y,z)$ is open. An analogous argument implies $A_\prec(x,y,z)$ is open.
\smallskip

\noindent $\mathbf{ \ref{it: open}\Rightarrow \ref{it: sarch}}$ Assume $\succcurlyeq$ is mixture-continuous and $A_\succ(x,z,y)$ and $A_\prec(x,z,y)$ are open for all $x,y,z.$  Pick $x,y,z\in \CM$ such that $x\succ y.$  Then, semi-transitivity and M1 imply $1\in A_\succ(x,z,y).$  It follows from $A_\succ(x,z,y)$ is open that there exists $\lambda<1$ such that $(\lambda, 1]\subset A_\succ(x,z,y).$ Similarly,  $1\in A_\prec(y,z,x).$ Since $A_\prec(y,z,x)$ is open, therefore there exists $\delta<1$ such that $(\delta, 1]\subset A_\prec(y,z,x).$  Hence, $\succcurlyeq$ is strongly Archimedean.

 The assertion $\mathbf{ \ref{it: sarch}\Rightarrow \ref{it: arch}}$ is immediate from the definition of strong Archimedean property. Therefore, the proof  of Proposition \ref{thm: arc} is complete. 
\prff


\prf[Proof of Proposition \ref{thm: mc}] 
Pick $x,y,z\in \CM.$ 
  If $A_\succcurlyeq(x,y,z)$ is empty, then it is closed. Then, assume $A_\succcurlyeq(x,y,z)\neq\emptyset$.  Since $A_\succcurlyeq(x,y,z)$  has finitely many components, it is the union of a finitely many non-empty, disjoint, convex sets $\{C_i\}_{i=1}^k$ which are closed in the subspace $A_\succcurlyeq(x,y,z)$.  Pick $i\leq k.$ Assume $\lambda=$ inf $C_i$ is not contained in  $C_i.$ Then, $C_i$ is a non-degenerate interval. It follows from $C_i$ is a component of $A_\succcurlyeq(x,y,z)$ that $x\lambda y\not\succcurlyeq z$. Since $A_{\bowtie}(x,y,z)$ is open and disjoint from $A_\succcurlyeq(x,y,z),$ therefore $\lambda\in A_\prec(x,y,z).$ Pick $\delta\in C_i.$ Then, $\delta>\lambda.$ It follows from $\succcurlyeq$ is strongly Archimedean that there exists $\beta\in (0,1)$ such that $z\succ (x\lambda y)\beta(x\delta y).$ Since $\succcurlyeq$ is semi-transitive and $(x\lambda y)\beta(x\delta y)\sim x(\beta\lambda + (1-\beta)\delta)y,$ therefore $z\succ x(\beta\lambda + (1-\beta)\delta)y.$ It follows from $\delta>\lambda$ that $\lambda<\beta\lambda + (1-\beta)\delta<\delta.$ Hence, $\beta\lambda + (1-\beta)\delta\in C_i.$ This furnishes us a contradiction. Therefore, $\lambda\in C_i$. An analogous argument shows that $C_i$ contains its supremum. Hence, $C_i$ is closed. As a union of finitely many closed sets, $A_\succcurlyeq(x,y,z)$  is closed.

An analogous argument shows that $A_\preccurlyeq(x,y,z)$ is closed. (Note that, for mixture sets, we do not need any transitivity property in order to prove this proposition.)
\prff


\prf[Proof of Theorem \ref{thm: ct}] 
Assume $\succcurlyeq$ is a non-trivial, reflexive, semi-transitive, mixture-continuous and Archimedean binary relation with a transitive symmetric part $\sim$ on a generalized mixture set $\CM.$  Recall that $\succ$ denote the asymmetric part of $\succcurlyeq.$ Fist consider the following claim. 

\cl
$\succ$ is negatively transitive. 
\lb{thm: negt}
\cll

It is easy to see that this claim implies $\succ$ is transitive. Then, it follows from the transitivity of $\sim$ and semi-transitivity of $\succcurlyeq$ that  $\succcurlyeq$ is transitive.\fn{See Khan-Uyan\i k \citeyearpar[Proposition 2]{ku17} for a proof and a detailed discussion on the interdependence between different transitivity conditions.} The following claim implies $\succcurlyeq$ is complete and transitive.

\cl
$\succcurlyeq$ is complete. 
\lb{thm: complete}
\cll

It remains to prove Claims \ref{thm: negt} and \ref{thm: complete} in order to complete the proof.

\prf[Proof of Claim \ref{thm: negt}]  Note that $\succ$ is negatively transitive if and only if $x\succ y$ implies $x\succ z$ or $z\succ y$ for all $x,y,z\in \CM.$ Assume $\succ$ is not negatively transitive, i.e. there exists $x,y,z\in \CM$ such that $x\succ y$ and neither $x\succ z$ nor $z\succ y.$

 It follows from semi-transitivity of $\succcurlyeq,$ transitivity of $\sim$ and M1  that $0\notin A_\sim(x,z,x)\cup A_\sim(y,z,y).$ Since $\succcurlyeq$ is reflexive, therefore M1 implies $1\in A_\sim(x,z,x)\cup A_\sim(y,z,y).$ Moreover, mixture-continuity implies $A_\sim(x,z,x)$ and $A_\sim(y,z,y)$ are closed subsets of [0,1], hence compact. Define $\lambda_x=\mmin A_\sim(x,z,x)$ and $\lambda_y=\mmin A_\sim(y,z,y).$ It is clear that $\lambda_x, \lambda_y \in (0,1].$ Define $\bar x=x\lambda_x z$ and $\bar y=y\lambda_y z.$ By construction, $\bar x\sim x$ and $\bar y\sim y,$ and it follows from semi-transitivity that  $\bar x\succ \bar y.$ 

  It follows from Proposition \ref{thm: arc} that $A_\prec(z,\bar y, \bar x)$ is open and from mixture-continuity that $A_\preccurlyeq(z,\bar y, \bar x)$ is closed. Since $\bar x\succ \bar y,$ therefore semi-transitivity of $\succcurlyeq,$ transitivity of $\sim$ and M1 imply $0\in A_\prec(z,\bar y, \bar x).$ It follows from semi-transitivity and M1 that $1\notin A_\prec(z,\bar y, \bar x).$  Since $A_\prec(z,\bar y, \bar x)$ is a non-empty strict subset of a connected set [0,1], therefore it cannot be both open and closed, hence $A_\prec(z,\bar y, \bar x)$ is not closed. Therefore, $A_\prec(z,\bar y, \bar x)\neq A_\preccurlyeq(z,\bar y, \bar x).$  This implies, there exists $\lambda_{\bar x}\in A_\preccurlyeq(z,\bar y, \bar x)\backslash A_\prec(z,\bar y, \bar x),$  i.e. $z \lambda_{\bar x} \bar y\sim \bar x.$ It is clear that $\lambda_{\bar x}\in (0,1].$ Define $\bar x_1=\bar y\left(1-\lambda_{\bar x}\right)z.$ It follows from M2 that $\bar x_1\sim  z\lambda_{\bar x} \bar y.$  An analogous argument implies there exists $\lambda_{\bar y}\in A_\preccurlyeq(z,\bar x, \bar y)\backslash A_\prec(z,\bar x, \bar y),$ i.e. $z \lambda_{\bar y} \bar x\sim \bar y.$ It is clear that $\lambda_{\bar y}\in (0,1].$ Define $\bar y_1=\bar x \left(1-\lambda_{\bar y}\right) z.$ It follows from M2 that $\bar y_1\sim z\lambda_{\bar y} \bar x.$  The transitivity of $\sim$ implies $\bar x_1\sim x$ and $\bar y_1\sim y.$ It follows from semi-transitivity that $\bar x_1\succ \bar y_1.$

Repeating the construction in the preceding paragraph one more time implies there exists $\lambda_{\bar x_1}\in (0,1]$ such that $z \lambda_{\bar x_1} \bar y_1\sim \bar x_1.$ Since $\sim$ is transitive, therefore  M2 implies $\bar y_1 \left(1-\lambda_{\bar x_1}\right) z\sim x.$ It follows from transitivity of $\sim$ and M3 that $\bar y_1 \left(1-\lambda_{\bar x_1}\right) z\sim x [(1-\lambda_{\bar x_1})(1-\lambda_{\bar y})\lambda_x] z,$ hence $(1-\lambda_{\bar x_1})(1-\lambda_{\bar y})\lambda_x\in A_\sim(x,z,x).$  Then $(1-\lambda_{\bar x_1})(1-\lambda_{\bar y})<1$ implies $\lambda_x\neq \mmin A_\sim(x,z,x)$ which furnishes us a contradiction. Therefore, $\succ$ is negatively transitive.
\prff


\prf[Proof of Claim \ref{thm: complete}] 
 Assume there exists $u,v\in \CM$ such that $u\bowtie v.$ It follows from non-triviality that $x\succ y$ for some $x, y\in \CM.$ Then, Claim \ref{thm: negt} implies  $x\succ u$ or $u\succ y$. 

Let $x\succ u.$ Then, Claim \ref{thm: negt} implies $x\succ v$ or $v\succ u$. Since $u\bowtie v,$ therefore $x\succ v.$ Hence, $x\succ u$ and $x\succ v.$ Next, we show that 
$$A_\succ(x,u,v)\cap A_\succ(x,u,u)=A_\succcurlyeq(x,u,v)\cap A_\succcurlyeq(x,u,u).$$ 
One of the inclusion relationship is trivial. In order to prove the other direction, pick $\lambda\in A_\succcurlyeq(x,u,v)\cap A_\succcurlyeq(x,u,u).$ If $x\lambda u\sim v,$ then it follows from transitivity of $\sim,$ semi-transitivity of $\succcurlyeq$ and  $x\lambda u\succcurlyeq u$ that $v\succcurlyeq u.$ This furnishes us a contradiction with $u\bowtie v.$ Hence, $x\lambda u\succ v,$ i.e. $\lambda\in A_\succ(x,u,v).$ Similarly, if $ x\lambda u\sim u,$ then $u\succcurlyeq v$ which contradicts $u\bowtie v.$ Hence, $\lambda\in A_\succ(x,u,u).$ 

It follows from M1, semi-transitivity and $u\bowtie v$ that $1\in A_\succ(x,u,v)\cap A_\succ(x,u,u)$ and $0\notin A_\succ(x,u,v)\cap A_\succ(x,u,u).$ Mixture-continuity imply $A_\succcurlyeq(x,u,v)\cap A_\succcurlyeq(x,u,u)$ is closed and Proposition \ref{thm: arc} imply $A_\succ(x,u,v)\cap A_\succ(x,u,u)$ is open. Therefore, we obtain a non-empty proper subset of [0,1] which is both open and closed. This furnishes us a contradiction with connectedness of [0,1].

The proof is analogous for $u\succ y.$ Therefore, $\succcurlyeq$ is complete.\fn{It is possible to apply the method of proof of Claim \ref{thm: negt} in order to prove this claim. The proof we provide here is simpler.}  
\prff

The proof of Theorem \ref{thm: ct} is complete.
\prff



\prf[Proof of Corollary \ref{thm: mixturest}]
If $\succcurlyeq$ is semi-transitive and $\sim$ is transitive, then applying Theorem \ref{thm: ct} finishes the proof. Therefore, it remains to prove $\succcurlyeq$ is semi-transitive and $\sim$ is transitive.\fn{This part of the proof does not use the properties of a generalized mixture set. Different versions are provided in \citet[Theorems 3 and 3$'$]{so65}, \citet[Theorem 1]{lo67} and \citet[Theorem I]{se69}. For completeness, we also provide a proof here.} First,  assume $\mathbf{\ref{it: rsemi}}$, i.e., $x\sim y\succ z$ implies $x\succ z$ for all $x,y,z\in \CM$.  In order to show that $\succcurlyeq$ is semi-transitive pick $x,y,z\in \CM$ such that $x\succ y\sim z.$  Assume $x\nsucc z.$ Then, completeness of $\succcurlyeq$ implies either $z\succ x$ or $z\sim x.$ If $z\succ x,$ then it follows from $y\sim z$ that $y\succ x,$ which furnishes us a contradiction. If $z\sim x,$ then $x\succ y$ implies $z\succ y$ which contradicts $z\sim y.$  Hence, $x\succ z$.  Therefore, $\succcurlyeq$ is semi-transitive. In order to show that  $\sim$ is transitive pick $x,y,z\in \CM$ such that $x\sim y\sim z.$ Assume $x\nsim z.$ Then completeness implies either $x\succ z$ or $z\succ x.$ Then, it follows from semi-transitivity that either $x\succ y$ or $z\succ y,$ which contradict $x\sim y$ and $y\sim z.$ Therefore, $\sim$ is transitive. An analogous argument shows that assertion $\mathbf{\ref{it: lsemi}}$ implies semi-transitivity of $\succcurlyeq$ and transitivity of $\sim$. 
\prff 


\prf[Proof of Theorem \ref{thm: mixturecat}]
Assume there exists $x,y,z\in\CM$ such that $x\succcurlyeq y\succcurlyeq z$ and $x\not\succcurlyeq z$. Then, completeness of $\succcurlyeq$ implies $z\succ x.$ It follows from $\succcurlyeq$ is strongly Archimedean that there exists $\lambda\in (0,1)$ such that $z\succ x\lambda y.$ Since $\succcurlyeq$ is star-convex and $\lambda\in (0,1),$ therefore $x\lambda y\succ y.$ Then, transitivity of $\succ$ implies $z\succ y$ which contradicts $y\succcurlyeq z.$ Therefore $\succcurlyeq$ is transitive.  An analogous argument proves the sufficiency of star-concavity for transitivity of $\succcurlyeq$. 
\prff 


The following result shows that convexity properties of preferences are characterized by the convexity of the certain subsets of [0,1]. 
\vspace{-5pt}

\lm
Let $\succcurlyeq$ be a reflexive binary relation with a transitive symmetric part $\sim$ on a mixture set $\CS.$ Then the following are valid.  
\ben[{\nf (a)}, topsep=3pt]
\setlength{\itemsep}{-1pt} 
\ml $\succcurlyeq$ is linear $\Leftrightarrow$ $A_\sim(x,y,z)$ is convex for all $x,y,z\in \CS$. \lb{it: ll}
\ml $\succcurlyeq$ is convex  $\Leftrightarrow$ $A_\succcurlyeq(x,y,z)$ is convex for all $x,y,z\in \CS$. \lb{it: cc}
\ml $\succcurlyeq$ is concave  $\Leftrightarrow$ $A_\preccurlyeq(x,y,z)$ is convex for all $x,y,z\in \CS$.  \lb{it: cvcv}
\ml Under mixture-continuity and Archimedean, $\succcurlyeq$ is linear $\Leftrightarrow$ it is  convex and concave.   \lb{it: lcc}
\een
\lb{thm: technical}
\lmm
\vspace{-10pt}

\prf[Proof of Lemma \ref{thm: technical}]  
\noindent $\mathbf{\ref{it: ll}}$  Assume $\succcurlyeq$ is linear.  Pick $\lambda, \delta\in A_\sim(x,y,z)$ and $\beta\in [0,1].$ Define $w_\lambda=x\lambda y$ and $w_\delta=x\delta y.$ By construction, $w_\lambda\sim z$ and $w_\delta\sim z.$ It follows from transitivity of $\sim$ that $w_\lambda\sim w_\delta.$ It follows from $\succcurlyeq$ is linear that $w_\lambda\sim w_\lambda \beta w_\delta.$ A simple algebra and S4 imply $w_\lambda \beta w_\delta=x(\beta\lambda+(1-\beta)\delta)y.$   It follows from transitivity of $\sim$  and $w_\lambda\sim z$ that $x(\beta\lambda+(1-\beta)\delta)y\sim z.$ Therefore, $\beta\lambda+(1-\beta)\delta\in A_\sim(x,y,z).$ Hence, $A_\sim(x,y,z)$ is convex.

Now assume $A_\sim(x,y,z)$ is convex for all $x,y,z\in \CS.$ Pick $x,y\in \CS$ and $\lambda\in [0,1]$ such that $x\sim y.$  It follows from reflexivity that $x\sim x.$ Therefore, S1 implies $0, 1\in A_\sim(x,y,x).$  Then the convexity assumption implies $A_\sim(x,y,x)=[0,1].$ Therefore, $\lambda, (1-\lambda)\in A_\sim(x,y,x).$ It follows from the transitivity of $\sim$ and $x\sim y$ that $x\lambda y\sim y$ and  $x(1-\lambda) y\sim x.$ Since $x(1-\lambda) y= y\lambda x,$ therefore $\succcurlyeq$ is linear. 
\medskip    


\noindent  $\mathbf{\ref{it: cc}}$ Assume $\succcurlyeq$ is convex.    Pick $\lambda, \delta\in A_\succcurlyeq(x,y,z)$ and $\beta\in [0,1].$ Define $w_\lambda=x\lambda y$ and $w_\delta=x\delta y.$ By construction, $w_\lambda\succcurlyeq z$ and $w_\delta\succcurlyeq z.$ It follows from $\succcurlyeq$ is convex  that $w_\lambda \beta w_\delta\succcurlyeq z.$ A simple algebra and S4 imply $w_\lambda \beta w_\delta=x(\beta\lambda+(1-\beta)\delta)y.$  Therefore, $\beta\lambda+(1-\beta)\delta\in A_\succcurlyeq(x,y,z).$ Hence, $A_\succcurlyeq(x,y,z)$ is convex.

Now assume $A_\succcurlyeq(x,y,z)$ is convex for all $x,y,z\in \CS.$  Pick $x,y,z\in \CS$ and $\lambda\in [0,1]$ such that $x\succcurlyeq z$ and $y\succcurlyeq z.$ It follows from S1 that $0,1\in A_\succcurlyeq(x,y,z).$ Since $A_\succcurlyeq(x,y,z)$ is convex, therefore  $A_\succcurlyeq(x,y,z)=[0,1].$ Hence, $x\lambda y\succcurlyeq z.$ Therefore,  $\succcurlyeq$ is convex. 
\medskip 


\noindent  $\mathbf{\ref{it: cvcv}}$ The proof is analogous to the proof of assertion \ref{it: cc}  above.  
\medskip 


\noindent   $\mathbf{\ref{it: lcc}}$ Assume $\succcurlyeq$ is linear.  Assertion \ref{it: ll}  above implies $A_\sim(x,y,z)$ is convex for all $x,y,z\in \CS,$ hence a connected subset of $[0,1].$ Note that $[0,1]\backslash A_\sim(x,y,z)=A_\succ(x,y,z)\cup A_\prec(x,y,z)\cup A_{\bowtie}(x,y,z).$ It is clear that the three sets are mutually disjoint. It follows from Proposition \ref{thm: arc} and mixture-continuity that they are open.\fn{Note that semi-transitivity is not needed in Proposition \ref{thm: arc} in a mixture set.}  By definition $A_\succcurlyeq(x,y,z)=A_\succ(x,y,z)\cup A_\sim(x,y,z).$ It follows from assertions \ref{it: cc}  and \ref{it: cvcv} above that convexity of $\succcurlyeq$ is equivalent to the convexity of $A_\succcurlyeq(x,y,z)$ and concavity of $\succcurlyeq$ is equivalent to the convexity of $A_\preccurlyeq(x,y,z)$  for all $x,y,z\in \CS.$ Therefore, the following result due to \citet[Theorem 9.9, p. 20]{wi49} completes the proof of this part.

\cl
If $C$ is a connected subset of a connected topological space $X$ such that $X\backslash C$ is the union of $n$ ($n>1$) non-empty, pairwise disjoint sets $A_i$ which are open in $X\backslash C,$ then $C\cup A_i$ is connected for all $i.$ 
\cll 
 
Now assume $A_\succcurlyeq(x,y,z)$ is convex and $A_\preccurlyeq(x,y,z)$  is concave for all $x,y,z\in \CS.$ By definition $A_\sim(x,y,z) = A_\succcurlyeq(x,y,z)\cap A_\preccurlyeq(x,y,z)$ for all $x,y,z\in \CS.$ Since intersection of two convex sets is convex, therefore $A_\sim(x,y,z)$ is convex for all $x,y,z\in \CS.$ Then, assertion \ref{it: ll}  above implies $\succcurlyeq$ is linear. (Note that we do not use the continuity assumptions in order to prove this direction of the assertion.)

Therefore, the proof of Lemma \ref{thm: technical} is complete.
\prff


\prf[Proof of Theorem \ref{thm: comparison}]
$\mathbf{\ref{it: linear}}$ Assume $\succcurlyeq$ is linear. Pick $x,y,z\in \CS$ such that $x\sim y$ and $y\succ z.$  Assume $x\nsucc z.$ If $x\sim z,$ then the transitivity of $\sim$ implies $y\sim z$ which contradicts $y\succ z.$ Therefore, either $z\succ x$ or $x\bowtie z.$

Proposition \ref{thm: arc} implies $A_\succ(x,y,z)$ and $A_\prec(x,y,z)$ are open. It follows from mixture-continuity that $A_{\bowtie}(x,y,z)$ is open. It is easy to see that the sets $A_\succ(x,y,z), A_\prec(x,y,z)$ and $A_{\bowtie}(x,y,z)$ are pairwise disjoint. It follows from $\sim$ is transitive, $x\sim y$ and $x\nsim z$ that $A_\sim(x,y,x)$ and $A_\sim(x,y,z)$ are disjoint. Therefore, 
\begin{equation}
\begin{array}{c}
A_\sim(x,y,x)=\left[ A_\succ(x,y,z) \cap A_\sim(x,y,x) \right] \cup \left[ \left(A_\prec(x,y,z)\cup A_{\bowtie}(x,y,z)\right) \cap A_\sim(x,y,x) \right] 
\end{array}
\lb{eq: asim}
\end{equation}

It is clear that the two sets in square brackets in Equation \ref{eq: asim} are pairwise disjoint and open in $A_\sim(x,y,x).$  Since $\succcurlyeq$ is reflexive and $x\sim y,$ therefore S1 implies $0, 1\in A_\sim(x,y,x).$ It follows from $y\succ z$ ans S1 that $0\in A_\succ(x,y,z).$ Therefore, $0\in A_\succ(x,y,z) \cap A_\sim(x,y,x).$ It follows from either $z\succ x$ or $x\bowtie z$, and S1 that either $1\in \left(A_\prec(x,y,z) \cup A_{\bowtie}(x,y,z)\right) \cap A_\sim(x,y,x).$ Therefore, $A_\sim(x,y,x)$ is the union of two non-empty, disjoint and open sets which contradicts Lemma \ref{thm: technical}\ref{it: ll} . Therefore, $x\succ z$. 

An analogous argument shows that $x\succ y$ and $y\sim z$ implies $x\succ z$ for all $x,y,z\in \CS$. Therefore, $\succcurlyeq$ is semi-transitive. 
\medskip


\noindent $\mathbf{\ref{it: convex}}$ Assume $\succcurlyeq$ is convex.  Pick $x,y,z\in \CS$ such that $x\sim y$ and $y\succ z.$  Assume $x\nsucc z.$ If $x\sim z,$ then the transitivity of $\sim$ implies $y\sim z$ which contradicts $y\succ z.$ Therefore, either $z\succ x$ or $x\bowtie z.$ 
  By definition
$$
[0,1]\backslash A_\sim(x,y,z)=A_\succ(x,y,z)\cup A_\prec(x,y,z)\cup A_{\bowtie}(x,y,z).
$$
It follows from Proposition \ref{thm: arc} and mixture-continuity that  $A_\succ(x,y,z), A_\prec(x,y,z)$ and $A_{\bowtie}(x,y,z)$ are open. It is clear that these three sets are pairwise disjoint.  Moreover, it follows from $y\succ z$ and S1 along with either $z\succ x$ or $x\bowtie z$ that $0\in A_\succ(x,y,z)$ and $1\in \left(A_\prec(x,y,z)\cup A_{\bowtie}(x,y,z)\right)$.  Therefore, there exists $\lambda'\in A_\sim(x,y,z),$ otherwise this yields a contradiction with the connectedness of $[0,1].$ It is clear that $\lambda'\in (0,1).$

It follows from S1, $x\sim y$ and reflexivity of $\succcurlyeq$ that $0, 1\in A_\succcurlyeq(x,y,y).$  Hence, Lemma \ref{thm: technical}\ref{it: cc}   implies $A_\succcurlyeq(x,y,y)=[0,1].$ Therefore, $ \lambda' \in A_\succcurlyeq(x,y,y).$ It follows from $y\succ z$ and $\sim$ is transitive that $ A_\sim(x,y,z)\cap A_\sim(x,y,y)=\emptyset.$ Hence, $\lambda'\in  A_\succ(x,y,y).$ 
 
   Define $z'=x\lambda' y.$ Then 
$$
[0,1]\backslash A_\sim(z',z,y)=A_\succ(z',z,y)\cup A_\prec(z',z,y)\cup A_{\bowtie}(z',z,y).
$$
It follows from S1, $z'\succ y$ and $y\succ z$ that $1\in A_\succ(z',z,y)$ and $0\in A_\prec(z',z,y).$  Analogous to the above argument, connectedness of $[0,1]$ implies there exists $\delta' \in A_\sim(z',z,y).$  It is clear that $\delta'\in (0,1).$

It follows from $\lambda'\in A_\sim(x,y,z)$ that $z'\sim z.$ Then,  S1 and reflexivity of $\succcurlyeq$ implies $0, 1\in A_\succcurlyeq(z',z,z').$ Hence, Lemma \ref{thm: technical}\ref{it: cc}  implies $A_\succcurlyeq(z',z,z')=[0,1].$ Therefore, $\delta' \in A_\succcurlyeq(z',z,z').$ It follows from $y\succ z$ and $\sim$ is transitive that $ A_\sim(z',z,y)\cap A_\sim(z',z,z')=\emptyset,$ therefore $\delta'\in  A_\succ(z',z,z').$ 

Define $y'=z'\delta' z.$ It follows from $\delta' \in A_\sim(z',z,y)$ and transitivity of $\sim$ that $y\sim y'$ and $x\sim y'.$ Then, it follows from Lemma \ref{thm: technical}\ref{it: cc} that $A_\succcurlyeq(x,y,y')=[0,1].$ Since $z'=x\lambda' y,$ therefore $z'\succcurlyeq y'.$ This furnishes us a contradiction with $\delta'\in  A_\succ(z',z,z'),$ i.e. $y'\succ z'.$ 
\medskip


\noindent $\mathbf{\ref{it: concave}}$ The proof is analogous to the proof of \ref{it: convex} above.  
\medskip 


\noindent $\mathbf{\ref{it: complete}}$ Let $\succcurlyeq$ be complete and convex. Then, it follows from assertion \ref{it: convex} above that $x\sim y$ and $y\succ z$ implies $x\succ z$ for all $x,y,z\in \CS.$ In order to show $\succcurlyeq$ is semi-transitive, pick $x,y,z\in \CS$ such that $x\succ y$ and $y\sim z.$ Assume $x\nsucc z.$ Then either $z\succ x$ or $z\sim x.$ If $z\succ x,$ the it follows from $y\sim z$ and convexity that $y\succ x,$ which furnishes us a contradiction. If $z\sim x,$ then convexity implies $z\succ y$ which contradicts $z\sim y.$  Hence, $x\succ z$.  Therefore, $\succcurlyeq$ is semi-transitive. The proof of the sufficiency of concavity for semi-transitivity is analogous. 

The proof of Theorem \ref{thm: comparison} is complete.
\prff


The proofs of Corollaries \ref{thm: mixture} and \ref{thm: mixturecc} are immediate from Theorems \ref{thm: ct} and \ref{thm: comparison}.


\prf[Proof of Theorem \ref{thm: realline}]
Assume $\succcurlyeq$ is a non-trivial, anti-symmetric, complete, transitive and mixture continuous binary relation on a generalized mixture set $\CM$ induced by $\sim$.  Since $\CM$ is induced by an anti-symmetric relation, it is a mixture set. For the convenience of reader, define $\CS=\CM$.  First, consider the following claim. 

\cl 
$\CS$ is isomorphic to a convex subset of some linear space. 
\lb{thm: isomorphism}
\cll

Assume without loss of generality that $\CS$ is a convex  subset of some linear space and $x\lambda y=\lambda x+(1-\lambda)y$ for all $x,y\in\CS$ and all $\lambda\in [0,1]$.  Pick $\bar{x},\bar{y}\in \CS$ such that $\bar{y}\succ \bar{x}$.\fn{Since $\succcurlyeq$ is non-trivial, there exist such pair.} We next show that for any $z\in \CS$, one and only one of the following is true.
\ben[{\nf (i)}, topsep=1pt]
\setlength{\itemsep}{-3pt} 
\ml there exists $\lambda\in (0,1)$ such that $\bar{x}=z \lambda \bar{y}$, \lb{it: xzy}
\ml there exists $\lambda\in [0,1]$ such that $z=\bar{x}\lambda \bar{y}$, \lb{it: zxy}
\ml  there exists $\lambda\in (0,1)$ such that $\bar{y}= \bar{x} \lambda z$. \lb{it: yxz}
\een
\nt  For any $z\in \CS$,  it follows from completeness and transitivity of $\succcurlyeq$ that one and only one of the following is true: (i$'$) $z\prec \bar{x}$, (ii$'$) $\bar{x}\preccurlyeq z\preccurlyeq \bar{y}$, (iii$'$) $\bar{y}\prec z$. If (i$'$) holds, then $z\prec \bar{x}\prec \bar{y}$. By mixture continuity and connectedness of $[0,1]$, there exists $\lambda\in (0,1)$ such that $z \lambda \bar{y} \sim \bar{x}$. Since $\succcurlyeq$ is anti-symmetric, $\bar{x}=z \lambda \bar{y}$.  Similarly, (ii$'$) implies that there exists $\lambda\in [0,1]$ such that $z=\bar{x}\lambda \bar{y}$, and (iii$'$) implies that there exists $\lambda\in (0,1)$ such that $\bar{y}= \bar{x} \lambda z$. Therefore, for any $z\in \CS$, at least one of (i), (iii), and (iii) holds. 
 In order to show that only one of them holds, pick $z\in \CS$. Assume (i) holds and $z\not\prec \bar{x}$, i.e., there exists $\lambda\in (0,1)$ such that $\bar{x}=z \lambda \bar{y}$ and $z\succcurlyeq \bar{x}$. Then, either $\bar{x}\preccurlyeq z\preccurlyeq \bar{y}$ or $\bar{x}\prec \bar{y}\prec z$, i.e., either (ii$'$) or (iii$'$) holds.  
 Assume (ii$'$) holds. Then, it follows from (i) that there exists $\lambda\in (0,1)$ such that $\bar x=\lambda z+ (1-\lambda) \bar y$. Then $\bar x\neq \bar y$ implies $\bar x\neq z\neq \bar y$. Since (ii$'$) implies (ii), therefore it follows from $\bar x\neq z\neq \bar y$ that there exists $\delta\in (0,1)$ such that $z=\delta \bar x+ (1-\delta) \bar y$. Then, $\bar x=\lambda \delta \bar x +(1-\lambda\delta)\bar y$ and $0<\lambda\delta<1$ furnish us a contradiction with $\bar x\neq \bar y$. Analogously, (iii$'$) yields a contradiction.  
 Therefore, $z\prec \bar{x}$, hence (i) and (i$'$) are equivalent.  Similarly,  (ii) implies $\bar{x}\preccurlyeq z\preccurlyeq \bar{y}$, and (iii) implies $\bar{y}\prec z$. Therefore, (ii) and (ii$'$), and (iii) and (iii$'$) are equivalent. 

 Therefore, the linear space containing $\CS$ is one dimensional. Without loss of generality, assume $\CS$ is an interval in $\Re$. 
 Since the usual orders $\geq$ and $\leq$ on $\Re$ are complete and anti-symmetric, therefore either $\bar x< \bar y$ or $\bar x > \bar y$. Assume $\bar x < \bar y$. Pick a pair $x,y\in \CS$ of distinct points. Assume without loss of generality that $x\prec y$. Then, one and only one of the following is true: (1) $y\prec \bar x$, (2) $y= \bar x$,  (3) $\bar x\prec y\prec \bar y$, (4) $y=\bar y$, (5) $\bar y\prec y$. If (1) holds, then it follows from $y\prec \bar x\prec \bar y$ and (i) above that there exists $\lambda\in (0,1)$ that $\bar x=\lambda y +(1-\lambda)\bar y$. Since $\bar x<\bar y$, therefore $y<\bar x$. Analogously, $x\prec y\prec \bar x$ and (i) imply $x<y$.  If (2) holds, then  it follows from $x\prec y=\bar x\prec \bar y$ and (i) that $x<\bar x=y$.  If (3) holds, then  it follows from $\bar x\prec y \prec \bar y$ and (i) that $y<\bar y$. Then, $x\prec y \prec \bar y$ and (i) imply $x<y$.  If (4) holds and $x\prec \bar x$, then  it follows from $x\prec \bar x\prec \bar y=y$ and (i) that $x<y$.  If (4) holds and $\bar x\prec x$, then  it follows from $\bar x\prec x\prec \bar y=y$ and (i) that $x<y$. The case where (4) holds and $x=\bar x$ trivially implies $x<y$.  Lastly, assume (5) holds. It follows from $\bar x\prec \bar y\prec y$ and (i) that $\bar x<\bar y< y$. If $x\prec \bar y$, then  it follows from $x\prec \bar y\prec y$, (i) and $\bar y<y$ that $x<y$. Similarly, if $\bar y\prec x$, then  it follows from $\bar y\prec x\prec y$, (i) and $\bar y<y$ that $x<y$. The case where $x=\bar y$ trivially implies $x<y$. Since $x,y$ are arbitrary, $x'\prec y'$ implies $x'<y'$ for all $x',y'\in \CS$. Conversely,  if $x<y$ and $x\nprec y$ for some $x,y\in \CS$, then anti-symmetry and completeness of $\preccurlyeq$ imply $y\prec x$. The argument above implies $y<x$ which furnishes us a contradiction. Then, $\prec$ and $<$ are identical. Since $\preccurlyeq$ is anti-symmetric, therefore $\preccurlyeq$ and $\leq$ are identical. If $\bar x > \bar y$, then an analogous argument implies that  $\preccurlyeq$ and $\geq$ are identical.  
   It remains to prove Claim \ref{thm: isomorphism}.

\prf[Proof of Claim \ref{thm: isomorphism}] 
 Proposition \ref{thm: shm} shows that if a mixture set $\CS$ satisfies conditions (C1) and (C2), then $\CS$ is isomorphic to a convex subset of some linear space. Therefore showing both conditions hold completes the proof. 

In order to show (C2), pick $x,y,z\in \CS$. If $x=y$ or $x=z$ or $y=z$, then S4 implies (C2). Thus, assume $x,y,z$ are distinct. Anti-symmetry of $\succcurlyeq$ implies that one and only one of the following cases holds: $x\succ y \succ z, x\succ z \succ y, y\succ x \succ z,y\succ z\succ x,z\succ x \succ y,z\succ y \succ x $. We only show (C2) holds in the first case since other cases can be proved similarly. In follows from  $x\succ y \succ z$ and (i) above that there exists $\alpha \in (0,1)$ such that $y=x\alpha z$. Then, S4 implies that for all $\lambda, \mu\in [0,1]$ with $\lambda\mu\neq 1$, 
\vspace{-12pt}

\begin{align}
(x\lambda y) \mu z&=\big(x\lambda (x\alpha z)\big) \mu z=\big(x(\lambda+(1-\lambda)\alpha)z\big)\mu z=x\big(\mu(\lambda+(1-\lambda)\alpha)\big)z,\notag\\
x (\lambda \mu) \big(y \frac{\mu(1-\lambda)}{1-\lambda\mu}z\big)&=x (\lambda \mu) \big((x\alpha z) \frac{\mu(1-\lambda)}{1-\lambda\mu}z\big)=x (\lambda \mu) \big(x \frac{\alpha\mu(1-\lambda)}{1-\lambda\mu}z\big)=x\big(\mu(\lambda+(1-\lambda)\alpha)\big)z. \notag
\end{align}
\vspace{-5pt}
 
\nt Thus, (C2) holds.

The following claim is useful in the remaining part of the proof.  
\cl 
If $x\neq y$, then $\mu\in (0,1)$ implies $x\neq x\mu y$. 
\lb{thm: equality}
\cll

\prf[Proof of Claim \ref{thm: equality}]
Assume $x\neq y$ and $x=x\mu y$ for some $\mu\in (0,1)$. Then, S3 implies $x=x\mu y=(x\mu y)\mu y=x\mu^2 y$. Repeating this argument $n$ times implies $y=y\mu^n x$ for any natural number $n$.
Since $x\neq y$, we have $x\succ y$ or $y \succ x$. If $x \succ y$, then mixture continuity implies $\{\lambda\in [0,1] : x\lambda y \prec x\}$ is an open set which contains 0. Thus, there exists $\delta>0$ such that, for any $\lambda \in [0,\delta)$, $ x\lambda y \prec  x$. Then, for big enough $n$, $\mu^n <\delta$. Hence,  $x=x\mu^n y \prec x$ furnishes us  a contradiction. Analogously, $y \succ x$ yields a contradiction. 
\prff

 In order to show (C1), we first show the following condition of \citet[(A0$'$)]{mo01} holds.

\ben[{\nf (C1$'$)}, topsep=1pt]
\ml {\it $\text{ If } x\neq y \text{ and } \lambda_1,\lambda_2\in [0,1], \text{ then } x\lambda_1 y= x\lambda_2 y \text{ implies } \lambda_1=\lambda_2$.}  \lb{it: c1p}
\een
%

\nt Pick $x,y\in \CS$ and $\lambda_1,\lambda_2\in [0,1]$ such that $x\neq y$ and $\lambda_1\neq \lambda_2$. Assume, without loss of generality, that $\lambda_1 >\lambda_2$. Assume $x\lambda_1 y=x\lambda_2 y$. First, let $\lambda_2=0$. Then, S1, S2 and $x\neq y$ imply $y=x\lambda_1 y$ and $\lambda_1 \in (0,1)$. Let $\mu =1-\lambda_1$. Then, S2 implies $y=y\mu x$, hence $\mu\in (0,1)$  yields a contradiction with Claim \ref{thm: equality}. 
 %
 Therefore, \ref{it: c1p} holds when $\lambda_2=0$.  
Now, let $\lambda_2>0$. If $\lambda_1=1$, then  S1, S2 and $x\neq y$ imply  $x=x\lambda_2 y$ and $\lambda_2 \in (0,1)$. This furnishes us a contradiction with Claim \ref{thm: equality}. 
 Now, assume $1>\lambda_1 >\lambda_2>0$. Since $ \left(1-\frac{\lambda_1 -\lambda_2}{ 1-\lambda_2}\right)(1-\lambda_2)=1-\lambda_1$, it follows from S2 and S3 that $x\lambda_1 y=x \frac{\lambda_1 -\lambda_2}{ 1-\lambda_2} (x\lambda_2 y)$.  
  Let $\mu=\frac{\lambda_1 -\lambda_2}{ 1-\lambda_2}$ and $z=x\lambda_2 y$. Since $\lambda_2 \in (0,1)$, Claim \ref{thm: equality} implies $z\neq x$. It follows from $x\lambda_1 y=x\lambda_2 y=z$ that $z=z(1-\mu)x$. Then, $\mu \in (0,1)$ yileds a contradiction with Claim \ref{thm: equality}. Therefore, proof of \ref{it: c1p} is complete. 

Next, we will  show that (C1) holds. Pick $x,y, y'\in \CS$ such that $y\neq y'$.  Assume $x\lambda y=x\lambda y'$ for some $x\in \CS$ and $\lambda\in (0,1)$. If $x=y$, then $x=x\lambda x=x\lambda y=x\lambda y'$ furnishes\fn{The fact that $x=x\lambda x$ follows from $ x=x1x =x0x=(x0x)\lambda x=(x1x)\lambda x=x\lambda x$; see \citet[Chapter 2]{fi82} for a detailed discussion on further implications of mixture set axioms.} us a contradiction with Claim \ref{thm: equality}. Similarly, $x=y'$ yields a contradiction. Hence, assume $y\neq x \neq y'$. Then, one and only one of the following cases holds: $x\prec y\prec y', x\prec y'\prec y',y\prec x\prec y',y\prec y'\prec x, y'\prec x\prec y,y'\prec y\prec x$. We provide proof of the first case, since other cases can be shown similarly.
Since $x\prec y\prec y'$, it follows from (i) above that there exists $\mu\in (0,1)$ such that $y=x\mu y'$. Thus, $x\lambda y'=x\lambda y=x\lambda(x\mu y')=x\big(\lambda+(1-\lambda)\mu\big)y'$. Then, \ref{it: c1p} implies $\lambda=\lambda+(1-\lambda)\mu$. It follows from $\lambda \in (0,1)$ that $\mu=0$. This furnishes us a contradiction. Therefore, (C1) holds.
\prff
The proof of Theorem \ref{thm: realline} is complete. 
\prff


\prf[Proof of Corollary \ref{thm: representation}] 
We first show that the mixture operation $[x]\lambda [y]=[x\lambda y]$ is well defined on the quotient set $\CM|\!\sim$. Since $\CM$ is a generalized mixture set, any element in the quotient set is assigned to a point in the set. For all $[x']=[x]$ and $[y']=[y]$, it follows from $x\sim x', y\sim y'$ and the independence axiom that $x\lambda y\sim x'\lambda y'$, hence $[x]\lambda [y]=[x\lambda y]=[x'\lambda y']=[x']\lambda [y']$. Hence, the mixture operation is a well-defined function. It follows from (i) $[x]1[y]=[x1y]=[x]$, (ii) $[x]\lambda [y]=[x\lambda y]=[y(1-\lambda x)]=[y](1-\lambda)[x]$ and (iii) $([x]\lambda[y])\delta[y]=[x\lambda y]\delta [y]=[(x\lambda y)\delta y]=[x(\lambda \delta) y]=[x](\lambda\delta)[y]$ that  the mixture operation defined above satisfies S1--S3, hence  $\CM|\!\sim$ is a mixture set. 

Completeness and transitivity of the derived relation $\hat \succcurlyeq$ on the quotient set $\CM|\!\sim$ follow from those of $\succcurlyeq$. In order to show that $\hat \succcurlyeq$ satisfies mixture continuity, pick $x, y,z\in \CM$. We will show that 
$$
A_\succcurlyeq(x,y,z)=A_{\hat \succcurlyeq}([x],[y],[z]).
$$ 
In order to show the forward direction, pick $\lambda$ such that $x\lambda y\succcurlyeq z$. It follows from transitivity of $\succcurlyeq$ that for all $w\sim x\lambda y$ and all $z'\sim z$, $w\succcurlyeq z'$. Hence, $[x]\lambda [y]=[x\lambda y]~\hat \succcurlyeq ~[z]$. In order to prove the backward direction, pick $\lambda$ such that $[x]\lambda [y] ~\hat\succcurlyeq ~[z]$. Then, $[x]\lambda [y]=[x\lambda y]$ and the definition of  $\hat\succcurlyeq$ imply $x\lambda y\succcurlyeq z$. Therefore, $\succcurlyeq$ is mixture continuous if and only if  $\hat\succcurlyeq$ is mixture continuous. 

Therefore, $\hat\succcurlyeq$ is complete, transitive, anti-symmetric and mixture continuous. Then, following Theorem \ref{thm: realline}, assume without loss of generality that the mixture set $\CM|\!\!\sim$ is an interval in the real line and $\hat\succcurlyeq$ is either $\geq$ or   $\leq$.  Let $\hat\succcurlyeq=\geq$. Define a utility function $\hat u:\CM|\!\sim\!\!\!\!~\ra \Re$ as $\hat u([x])=[x]$. Note that $\hat u([x]\lambda[y])=[x]\lambda [y]=\lambda \hat u([x])+(1-\lambda)\hat u([y])$. Define $u:\CM\ra \Re$ as $u(x)=\hat u([x]).$ Since $u(x\lambda y)=\hat u([x\lambda y])=\hat u([x]\lambda [y])=\lambda \hat u([x])+(1-\lambda)\hat u([y])=\lambda u(x)+(1-\lambda) u(y)$, therefore $u$ has the desired properties for an expected utility representation of $\succcurlyeq$. If $\hat\succcurlyeq=\leq$, then the argument above with $\hat u([x])=-[x]$ completes the proof. 
\prff


\prf[Proof of Proposition \ref{thm: fragile}] It follows from Theorem \ref{thm: ct} that $\succcurlyeq$ is not Archimedean. Then Proposition \ref{thm: arc} implies there exists $x,y,z\in \CM$ such that $A_\succ(x,y,z)$ or $A_\prec(x,y,z)$ is not open. Assume without loss of generality that $A_\succ(x,y,z)$  is not open. Then there exists $\lambda\in A_\succ(x,y,z)$ and a sequence $\{\lambda_t\}_{t\in \mathbb N}$ such that $\lambda_t\rightarrow \lambda$ and $x  \lambda_t y \nsucc z$ for $t$  large enough. 

Assume $z \succcurlyeq x  \lambda_t y$ for $t$ large enough. Then, it follows from mixture-continuity that $A_\preccurlyeq(x,y,z)$ is closed, hence $\lambda\in A_\preccurlyeq(x,y,z).$ This furnishes us a contradiction with $x\lambda y\succ z.$ Therefore, $\{\lambda_t\}$ has a subsequence $\{\delta_t\}$ such that $\delta_t\in A_{\bowtie}(x,y,z)$ for $t$ large enough. It follows from mixture-continuity that $A_{\bowtie}(x,y,z)$ is open. Therefore for all $t$ large enough, $A_{\bowtie}(x,y,z)$ contains an open neighborhood $V_{\delta_t}$ of $\delta_t$.  Pick an open neighborhood $V_\lambda$ of $\lambda.$  It follows from $\delta_t\rightarrow \lambda$ that $\delta_t$ is contained in $V_\lambda$ for $t$ large enough. Therefore, the non-empty open set $V_\lambda\cap V_{\delta_t}$ is contained in $A_{\bowtie}(x,y,z)$ for $t$ large enough.
\prff

%
%
%
%

\prf[Proof of Proposition \ref{thm: flimsy}]
Assume $\succcurlyeq$ satisfies the hypotheses of the proposition and it is not flimsy. Then, for all  $x,y,z\in \CM$ and all $\lambda \in A_{\bowtie}(x,y,z)$, there exists an open neighborhood $V_\lambda$ of $\lambda$ such that $V_\lambda \subset  A_{\bowtie}(x,y,z)$. Then, $A_{\bowtie}(x,y,z)$ is open, and hence it has open sections. Then, it follows from Proposition \ref{thm: mc} that $\succcurlyeq$ is mixture-continuous. Then, Theorem \ref{thm: ct} implies $\succcurlyeq$ is complete. This furnishes us a contradiction. 
 \prff



\section{An Appendix of Examples}

In this section,  we provide some examples in mixture sets illustrating how Propositions  \ref{thm: arc}  and  \ref{thm: mc}  fail if any of its assumptions is dropped. Let $[0,1]$ be the choice space endowed with the usual structures.  It is clear that $[0,1]$ is a mixture set with the usual linear space operations. The following example concerns Proposition \ref{thm: arc}. 
 
\exm{\nf 
Assume $\succcurlyeq$ is a reflexive relation on [0,1] such that $1\succ x$ for all $x<1$ and no other elements are comparable. Then, each of $A_\succcurlyeq(x,y,z)$ and $A_\preccurlyeq(x,y,z)$  either contains at most two elements or equals [0,1] for all $x,y,z\in [0,1]$, hence both are closed. However, $A_\succ(1,0,0)=\{1\}$ is not open. Moreover, $1\succ 0$, but for all $x\in (0,1)$ and all $\lambda \in (0,1)$, $1\lambda x\nsucc 0$. Therefore, $\succcurlyeq$ is mixture-continuous, but violates any of the three equivalent conditions of Proposition \ref{thm: arc}. 
}\lb{ex: archmc1}
\exmm


 We next provide examples concerning Proposition \ref{thm: mc}.
\vspace{-3pt}

\exm{\nf  First, we illustrate $A_{\bowtie}(x,y,z)$ is not open for some $x,y,z.$ Define a binary relation $\succcurlyeq$ as follows: $x\sim y$ for all $x,y\in [0,0.5)$ and $x\bowtie y$ for all other points in [0,1]. Then $A_{\bowtie}(1,0,0)=[0.5, 1]$ which is not open.  Since $A_\succcurlyeq(x,y,z)$ and $A_\preccurlyeq(x,y,z)$ are equal to one of $\emptyset$ and an interval in [0,1] for all $x,y,z\in [0,1],$  therefore they are convex. 
 Strong Archimedean property is trivially satisfied since there is no $x,y$ such that $x\succ y.$ It follows from $A_\succcurlyeq(0,1,0)=[0,0.5)$ that $\succcurlyeq$ is not mixture-continuous.  
}\lb{ex: mcarch1}
\exmm

\exm{\nf  Second, we illustrate $\succcurlyeq$ is not strongly Archimedean. Define a binary relation $\succcurlyeq$ as follows: $x\sim y$ for all $x,y\in [0,0.5)$ and all $x,y\in [0.5,1]$, and $x\succ y$ for all $x\in [0.5,1], y\in [0,0.5).$ Then, $\succcurlyeq$ is complete.  It follows from $0.5\succ 0$ and $0.5\lambda 0\sim 0$ for all $\lambda\in (0,1)$ that $\succcurlyeq$ is not strongly Archimedean. Since $A_\succcurlyeq(x,y,z)$ and $A_\preccurlyeq(x,y,z)$ are equal to an interval in [0,1] for all $x,y,z\in [0,1],$  therefore they are convex. Observing that $A_\preccurlyeq(0,1,0)=[0,0.5)$ is not closed shows that $\succcurlyeq$ is not mixture-continuous.  
}\lb{ex: mcarch2}
\exmm

\exm{\nf  Third, we illustrate $A_\succcurlyeq(x,y,z)$ and $A_\preccurlyeq(x,y,z)$ do not have finitely many components. Define a binary relation $\succcurlyeq$ as follows. Assume $x\sim y$ if $x,y$ rational or $x,y$ irrational. Assume $x\succ y$ for all $x$ rational and $y$ irrational. For any rational $z$, $A_\succcurlyeq(1,0,z)=\mathbb Q\cap [0,1]$ and for any irrational $z'$, $A_\preccurlyeq(1,0,z')=(R\backslash Q)\cap[0,1]$. Both sets have infinitely many components. Moreover, $A_{\bowtie}(x,y,z)$ is empty for all $x,y,z,$ hence open. It follows from the ``Archimedean'' property in mathematics that $\succcurlyeq$ is strongly Archimedean. In order to see this, pick $x,y,z$ such that $x\succ y.$ Then, it is clear that $x$ is rational and $y$ is irrational. If $z$ is rational, then it follows from $x\lambda z$ is rational for all rational $\lambda\in [0,1]$ that $x\lambda z\succ y.$ Similarly, since $y\lambda z$ is irrational for all rational $\lambda\in (0,1),$ therefore $x\succ y\lambda z.$  If $z$ is irrational, then either $x<z$ or $z<x.$ In both cases the open interval determined by $x$ and $z$ contains a rational number $x'$ such that $x'=x\lambda z$ for some $\lambda\in [0,1].$ Hence $x\lambda z\succ y.$ Similarly, if $z$ is irrational and $z=y,$ then $x\succ y\lambda z$ for all $\lambda\in [0,1]$. Assume $z\neq y.$ Then the open interval determined by $y$ and $z$ contains an irrational number $y'$ such that $y'=y\lambda z.$ Hence $x\succ y\lambda z.$ Therefore, $\succcurlyeq$ is strongly Archimedean.   
 It follows from the set of rationals is not closed and that $A_\succcurlyeq (1,0,z)$ is the set of rationals in [0,1] for any rational $z$ that $\succcurlyeq$ is not mixture-continuous. 
}\lb{ex: mcarch3}
\exmm

 In the last example, both  $A_\succcurlyeq(x,y,z)$ and $A_\preccurlyeq(x,y,z)$ do not have finitely many components. We do not have an example where only one of these fail.  Hence it is possible to weaken this assumption, but we do not know such weaker assumption at present. It is easy to show that under independence assumption,  both  $A_\succcurlyeq(x,y,z)$ and $A_\preccurlyeq(x,y,z)$ are convex, hence have finitely many components. We show in Lemma \ref{thm: technical} in Section \ref{sec: proofs} that the convexity of these sets follows from linearity under mixture-continuity and Archimedean properties.

\medskip

\setlength{\bibsep}{4.1pt}
\setstretch{1}

%
\bibliographystyle{/Users/METIN/Dropbox/Research/SourceFiles/econometrica} 
\bibliography{/Users/METIN/Dropbox/Research/SourceFiles/References.bib}

\end{document}